\documentclass[notitlepage,nofootinbib,aps,prd,10pt,twocolumn]{revtex4-1}

\usepackage{amssymb}
\usepackage{physics}
\usepackage{graphicx}
\usepackage{comment}

\usepackage[utf8]{inputenc}
\usepackage{color}

\usepackage{enumitem}
\newlist{myitemize}{enumerate}{10}
\setlist[myitemize]{label*=\arabic*.,nosep,leftmargin=*}

\newcommand{\Pobs}{P_{\mathrm{obs}}}
\newcommand{\lmax}{\ell_{\mathrm{max}}}

\newcommand\lcdm{$\Lambda$CDM}

\newcommand{\beq}{\begin{equation}}
\newcommand{\eeq}{\end{equation}}
\newcommand{\beqa}{\begin{eqnarray}}
\newcommand{\eeqa}{\end{eqnarray}}

\begin{document}

\title{Higgs-dilaton cosmology: \\An inflation -- dark-energy connection and forecasts for future galaxy surveys}

\author{Santiago Casas}
\author{Martin Pauly}
\author{Javier Rubio}
\affiliation{Institut f\"ur Theoretische Physik, Ruprecht-Karls-Universit\"at Heidelberg, \\
Philosophenweg 16, 69120 Heidelberg, Germany}

\begin{abstract}
The Higgs-Dilaton model is a scale-invariant extension of the Standard Model non-minimally coupled to gravity and  
containing just one additional degree of freedom on top of the Standard Model particle content. This minimalistic scenario 
predicts a set of measurable consistency relations between the inflationary observables and the dark-energy 
equation-of-state parameter. We present an alternative derivation of these consistency relations that highlights 
the connections and differences with the $\alpha$-attractor scenario. We study how far these constraints allow one to 
distinguish the Higgs-Dilaton model from \lcdm{} and $w$CDM cosmologies. To this end we first analyze existing data sets 
using a Markov Chain Monte Carlo approach. Second, we perform forecasts for future galaxy surveys using a Fisher 
matrix approach, both for galaxy clustering and weak lensing probes. Assuming that the best fit values in the 
different models remain comparable to the present ones, we show that both Euclid- and SKA2-like missions will be 
able to discriminate a Higgs-Dilaton cosmology from $\Lambda$CDM and $w$CDM.
\end{abstract}

\maketitle

\section{Introduction} 

We are entering an era of precision cosmology in which many of the ingredients of our standard cosmological scenario   
will be tested with an unprecedented level of precision. By ruling out a vast set of models, 
future cosmological surveys such as DESI, Euclid  or SKA will shed light on fundamental aspects of modern physics such 
as inflation and dark energy.

The interpretation of future cosmological observations is inevitably influenced by our initial set of assumptions. An example of 
this is the usual treatment of inflation and dark energy as two completely independent 
epochs in the history of the Universe. Note, however, that there is no fundamental reason for this to be the case. 
Inflation and 
dark energy indeed share many essential properties that could be related to some 
underlying principle able to unify them within a common framework. Among the different implementations of this idea
proposed in the 
literature, the models based on scale and conformal symmetry are particularly interesting 
\cite{Wetterich:1987fm,Wetterich:1987fk,Wetterich:1994bg,Meissner:2006zh,Shaposhnikov:2008xb,GarciaBellido:2011de,Blas:2011ac,
Bezrukov:2012hx,Khoze:2013uia,Kannike:2015apa,Ferreira:2016vsc,Ferreira:2016wem,Kannike:2016wuy,Shkerin:2016ssc,Rubio:2017gty} since they could additionally
alleviate the Standard Model hierarchy problem \cite{Bardeen:1995kv,Shaposhnikov:2008xi,Tavares:2013dga} 
(see also Refs.~\cite{Karananas:2016grc,Karananas:2017zrg}). 

In this paper we will assume scale invariance to be an exact but spontaneously broken 
symmetry \cite{Shaposhnikov:2008xi}. The simplest 
realization of this idea is the so-called Higgs-Dilaton (HD) scenario 
\cite{Shaposhnikov:2008xb,GarciaBellido:2011de}. 
This model is a minimalistic extension of the Standard Model (SM) based on dilatation symmetry 
and unimodular gravity. Both the Planck scale and the Higgs vacuum expectation value are replaced 
by a singlet scalar field $\chi$, which, together with the Higgs field $H$, is allowed to 
be non-minimally coupled to gravity. While the non-minimal couplings allow for inflation with the usual
SM Higgs potential, the unimodular restriction allows to recover the late time 
acceleration of the Universe \textit{even in the absence of a cosmological constant} in the action. 
The HD model is close to \lcdm{} \,in terms of background  evolution  but contrary to this scenario it connects
the early and the late Universe in a very non-trivial way. In particular, it relates the existence of a dynamical 
dark energy component to deviations from the $\alpha$-attractor inflationary predictions 
\cite{Ferrara:2013rsa,Kallosh:2013yoa,Galante:2014ifa}. This is a unique connection between 
two eras far apart in the evolution of the Universe that could be potentially tested with future cosmological 
surveys.

In this work we study the impact of the HD consistency relations on cosmological observables. The main 
ideas and phenomenology of the HD model are introduced in Section \ref{sec:HDM}, 
where we present a novel derivation of the HD consistency relations that highlights 
the connections and differences with the well-known $\alpha$-attractor scenario 
\cite{Ferrara:2013rsa,Kallosh:2013yoa,Galante:2014ifa}. In Section \ref{sec:data1} we use
a Markov Chain Monte Carlo approach to analyze current data sets in the presence of the HD constraints.  
We discuss the chances of differentiating the HD model from other cosmological scenarios such as $\Lambda$CDM or $w$CDM. 
The impact  of the consistency relations on the results of future 
cosmological observations is explored in Section \ref{sec:forecast} via  a Fisher matrix approach. 
Our reference surveys are a DESI-like \cite{desi_collaboration_desi_2016-1,desi_collaboration_desi_2016}, 
a Euclid-like \cite{amendola_cosmology_2013, laureijs_euclid_2011}
and a SKA2-like \cite{yahya_cosmological_2015,santos_hi_2015,raccanelli_measuring_2015,bull_measuring_2015} galaxy survey 
measuring the expansion history of the Universe and the evolution
of large scale structures up to a redshift $z\sim 2$. The conclusions of our analysis are 
presented in Section \ref{sec:conclusions}.

\section{The Higgs-Dilaton model}\label{sec:HDM}

The  key ingredients of the HD model are scale-invariance (SI) and unimodular gravity (UG). In the unitary gauge $H=(0,h/\sqrt{2})^T$,
the graviscalar sector of the HD Lagrangian density takes the form \cite{Shaposhnikov:2008xb,GarciaBellido:2011de} 
\beq
\label{eqn:hdm_lagrangian}
\frac{{\cal L}_{SI+UG}}{\sqrt{-g}} = \frac{f(h,\chi)}{2}
R 
-\frac12 (\partial h)^2- \frac{1}{2} (\partial \chi)^2-V(h,\chi)\,,
\eeq
with 
\beq\label{eq:f}
f(h,\chi)=\xi_h h^2 +\xi_\chi \chi^2 \,,
\eeq 
and 
\beq \label{eq:pot}
 \hspace{5mm}V(h,\chi)=\frac{\lambda}{4} \left(h^2-\alpha\chi^2 \right)^2 + \beta \chi^4
\eeq
a scale-invariant potential with $\alpha\geq 0$, $\beta\geq 0$. In order to have a well-defined 
graviton propagator for all field values, the non-minimal couplings  
$\xi_h$ and $\xi_\chi$ must be positive definite, $\xi_h,\xi_\chi>0$. 
The Lagrangian density \eqref{eqn:hdm_lagrangian} is supplemented with the SM Lagrangian density
without the Higgs potential, ${\cal L}_{{\rm SM}[\lambda\to 0]}$. Apart from a potential dark matter candidate, 
no additional degrees of freedom are added on top of this minimalistic matter content. 

In the absence of gravity, the ground state of the Higgs-Dilaton system corresponds to the minima 
of the scale-invariant potential \eqref{eq:pot}. For $\alpha\neq 0$ and $\beta=0$, this potential 
contains two flat directions with 
\beq \label{flatd}
h_0^2=\alpha \chi_0^2\,,
\eeq and arbitrary 
$\chi_0$ that lead to the spontaneous symmetry breaking of scale invariance. The case $\beta\neq 0$  
translates into a physically unacceptable ground state, $h_0=\chi_0=0$, containing a massless Higgs boson and no 
electroweak symmetry breaking.

The inclusion of gravity via the non-minimal couplings $\xi_h$ and $\xi_\chi$ 
results in the appearance of an additional ground state 
\beq 
h_0^2=\alpha \chi_0^2+\frac{\xi_h}{\lambda} R\,.
\eeq  
Depending on the value of $\beta$, the 
background spacetime corresponds to a flat ($\beta=0$), de Sitter ($\beta>0$) or anti de Sitter ($\beta<0$) geometry,
\beq 
R=\frac{4\beta\chi_0^2}{\xi_\chi+\alpha \xi_h}\,.
\eeq  
Among
the possible values  of $\beta$, the case $\beta=0$ seems to be preferred from the quantum 
field theory point of view. Indeed, the presence of the massless scalar field $\chi$ in both de Sitter and 
anti de Sitter backgrounds is known to give rise to instabilities \cite{Allen:1987tz,Jalmuzna:2011qw}, 
see also Refs.~\cite{Antoniadis:1985pj,Tsamis:1992sx,Tsamis:1994ca,Antoniadis:2006wq,Polyakov:2009nq}. These results seem to be in conceptual 
agreement with those following from functional renormalization group approaches involving non-minimally coupled 
scalar fields \cite{Wetterich:2017ixo}.
On top of that, the case $\beta=0$ allows for the spontaneous symmetry breaking of scale 
invariance \textit{even in the absence of gravity}. Based on these arguments, we will focus on the 
$\beta=0$ case in what follows. 

In unimodular gravity \cite{vanderBij:1981ym,Wilczek:1983as,Buchmuller:1988wx,Unruh:1988in,Weinberg:1988cp,Henneaux:1989zc,Buchmuller:1988yn} 
the metric determinant in Eq.~\eqref{eqn:hdm_lagrangian} is restricted to take a constant value $g = -1$. As General Relativity, UG 
can be understood as a particular case of a theory invariant under transverse diffeomorphisms,
$x^\mu\to x^\mu+\xi^\mu$ with $\partial_\mu \xi^\mu=0$, in which 
 the third metric degree of freedom is absent  \cite{Buchmuller:1988wx,Pirogov:2006zd,Alvarez:2006uu}. Note that 
 the unimodular constraint is not a strong restriction  since the resulting theory of gravity can still describe all possible geometries.

 The presence 
 of the unimodular constraint $g = -1$ translates into the appearance of an integration constant ${\Lambda_0}$ at the level 
 of the equations of motion  \cite{Shaposhnikov:2008xb}. These equations coincide with those obtained from 
 a \textit{diffeomorphism invariant} Lagrangian \cite{Shaposhnikov:2008xb}
\beq\label{eqn:hdm_lagrangianL}
\frac{\cal L}{\sqrt{-g}}=\frac{f(h,\chi)}{2} R  -\frac{1}{2} (\partial h)^2 -\frac{1}{2} (\partial \chi)^2-V(h,\chi)+{\Lambda_0}\,,
\eeq
but with ${\Lambda_0}$ understood as an initial condition rather than a cosmological constant. Given the equivalence 
of the two formulations, we will work with the more familiar diffeomorphism invariant Lagrangian
\eqref{eqn:hdm_lagrangianL}, but keeping in mind the aforementioned interpretation of $\Lambda_0$.

\subsection{Higgs-Dilaton Cosmology} 

As explained in the previous section any solution with $\beta=0$ and $\chi_0\neq 0$ leads to the spontaneous symmetry 
breaking of scale invariance. The expectation value of the field $\chi$ induces the masses of the SM particles through its coupling to the Higgs field 
and defines the effective Planck mass. The hierarchy between the Planck mass $M_P$ and 
the electroweak scale $v$ is reproduced by  properly fine-tuning the $\alpha$ coupling to a value 
$\sim v^2/M_P^2\sim 10^{-32}$ \cite{Shaposhnikov:2008xb,GarciaBellido:2011de}. Due to the small value of this coupling constant,
the flat directions in Eq.~\eqref{flatd} are essentially aligned with the direction of the $\chi$ field. For all practical purposes 
it will be enough to consider the Lagrangian density \eqref{eqn:hdm_lagrangianL} in the $\alpha\simeq 0$ approximation, but 
keeping in mind that the potential really contains two almost-degenerate valleys. 

The cosmological consequences of the HD model are 
more easily understood in the so-called Einstein frame, in which the gravitational part of the action takes the usual 
Einstein-Hilbert form. Performing a metric rescaling $  g_{\mu\nu} \rightarrow \Omega^{-2} g_{\mu\nu}$ with conformal factor 
\beq \label{eq:conf}
\Omega^2 = f(h,\chi)/M_P^{2}\,,
\eeq 
we obtain the Einstein-frame Lagrangian density
\beq\label{SE1}
\frac{\cal L}{\sqrt{-g}}=\frac{M_P^2}{2} R - 
\frac{1}{2} g^{\mu\nu} \gamma_{ab} \partial_\mu \varphi^a \partial_\nu \varphi^b - U_{\rm E}(\varphi^a)\,,
\eeq
where we have made use of a compact notation $\varphi^{a}=(\varphi^1,\varphi^2)=(h,\chi)$ to define a field-space metric
\beq
  \gamma_{ab} = \frac{1}{\Omega^2}\left( \delta_{ab} + \frac{3}{2} M_P^2 
  \frac{\partial_a \Omega^2 \partial_b \Omega^2}{\Omega^2}\right)\,.
\eeq 
\begin{figure}
 \includegraphics[width=0.4\textwidth]{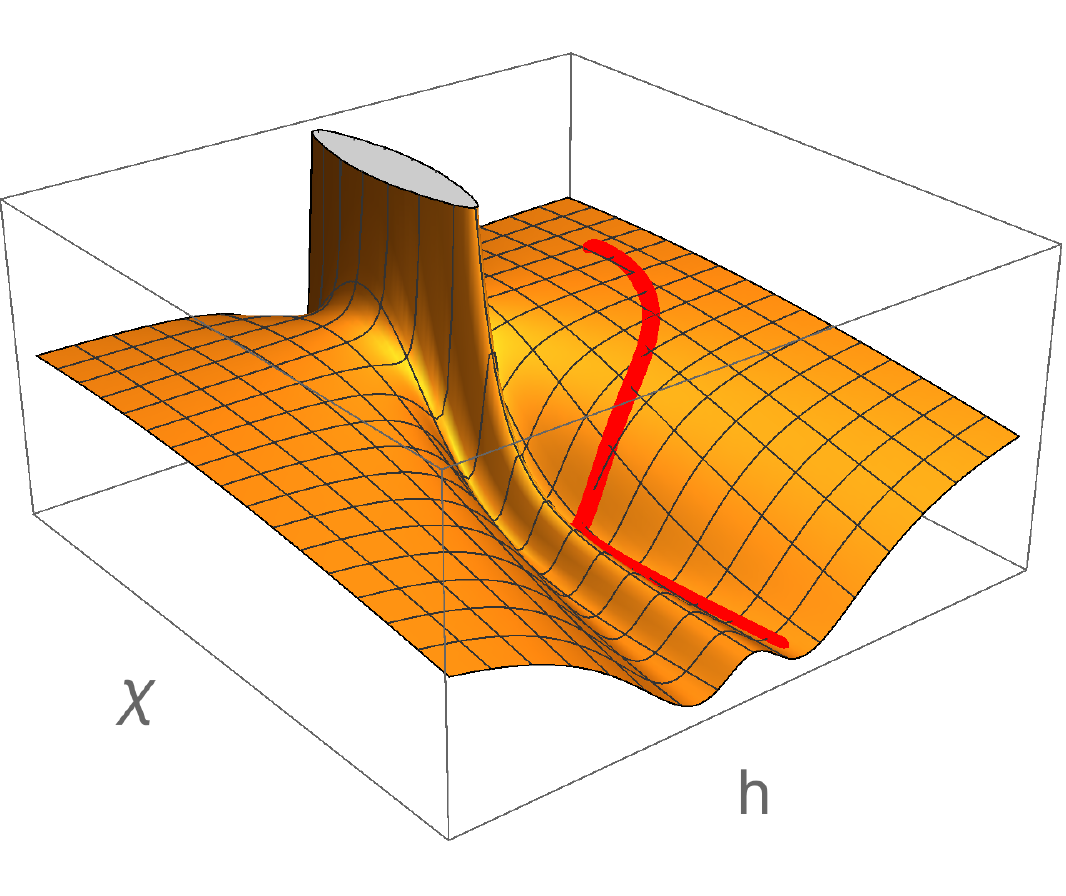}
 \caption{\label{fig:potential} The HD Einstein-frame potential $U_E=U+U_{\Lambda_0}$ in the $\Lambda_0>0$ case. In order 
 to better visualize the almost degenerate valleys at $h_0=\pm \sqrt{\alpha} \chi_0$ we used a rather large and unphysical value of $\alpha$.
 For illustration purposes we also included a typical Higgs-Dilaton trajectory. Inflation takes place in the asymptotically 
 flat region, where the effective potential can be well approximated by the $U$ term and scale-invariance 
 is approximately realized. The conservation of the Noether current associated to this continuous symmetry restricts the motion 
 of the fields to ellipsoidal trajectories in the $\lbrace h,\chi\rbrace$ plane. After inflation, heating and entropy production take 
 place and the fields eventually relax to the valleys of the potential. Due to the $U_{\Lambda_0}$ term 
 in Eq.~\eqref{eq:potUdef}, these valleys are slightly tilted towards $\chi\rightarrow\infty$. At early times, the 
 Hubble friction is so large that the fields stay essentially ``frozen''. When the decreasing SM energy 
 density equals the approximately constant term $U_{\Lambda_0}$, the fields roll down 
 the valley and asymptotically approach the ground state of the system at $\chi\rightarrow\infty$.}
\end{figure}
The Einstein-frame potential $U_{\rm E}=U+U_{\Lambda_0}$ is given by the sum of two pieces,
\beq\label{eq:potUdef}
U(h,\chi) = \frac{\lambda}{4}\frac{M_P^4 \,h^4}{f^2(h,\chi)} \,,\hspace{5mm} 
U_{\Lambda_0}(h,\chi) = \frac{{\Lambda_0} M_P^4}{f^2(h,\chi)}\,,
\eeq
whose effect is illustrated in Fig.~\ref{fig:potential}. The term $U_{\Lambda_0}$ modifies the flat directions \eqref{flatd} associated to 
the spontaneous symmetry breaking of scale invariance. For $\Lambda_0<0$ the vacuum valleys become tilted towards the 
origin, leading to a trivial ground state with $\chi=h=0$ and no particle excitations. For $\Lambda_0>0$, the 
ground state of the system  is rather located at $\chi\rightarrow\infty$. Far away from the symmetry-breaking 
valleys, the contribution $U_{\Lambda_0}$ becomes negligible and the effective potential $U_E $ can be well approximated by the 
$U$ term, which becomes asymptotically flat for sufficiently large values of the Higgs field.

The non-diagonal metric in Eq.~\eqref{SE1} can be recast into a diagonal form by considering an additional field redefinition 
$\lbrace h,\chi\rbrace\to\lbrace \Theta,\Phi\rbrace$ with 
\beqa
\gamma^{-2} \Theta &\equiv& \frac{(1+6\xi_h)h^2+(1+6\xi_\chi)\chi^2}{\xi_h h^2+\xi_\chi\chi^2}\,,\label{eq:Thetadef} \\
\exp\left[\frac{2\gamma\Phi}{M_P}\right] &\equiv& \frac{\kappa_c}{\kappa} \frac{(1+6\xi_h)h^2+(1+6\xi_\chi)\chi^2}{M_P^2}\,.
\eeqa
For future convenience we have also defined several non-minimal coupling combinations, namely 
\beq\label{kappadef}
\kappa_c \equiv-\frac{\xi_h}{1+6\xi_h}\,, \hspace{10mm} 
\kappa \equiv\kappa_c\left(1-\frac{\xi_\chi}{\xi_h}\right)\,,
\eeq
and
\beq
\gamma \equiv \sqrt{\frac{\xi_\chi}{1+6\xi_\chi}}\,.
\eeq 
Note that the variable $\Theta$ in Eq.~\eqref{eq:Thetadef} is a function of the ratio of $h$ to $\chi$ and thus it does not transform 
under scale transformations. This type of transformations acts only on the $\Phi$ field, which can be interpreted 
as a generalized radial coordinate in the $\lbrace h,\chi\rbrace$ plane. This variable is defined in such a way that 
scale transformations act on it as shift transformations. In terms 
of the coordinates $\lbrace\Theta,\Phi\rbrace$, the Lagrangian density \eqref{SE1} becomes
\begin{equation}\label{action_HD2}
\frac{\cal L}{\sqrt{-g}}=  \frac{M_P^2}{2}R 
-\frac{K(\Theta)}{2}(\partial\Theta)^{2} 
- \frac{\Theta}{2}(\partial \Phi)^2 
 -U-U_{\Lambda_0}\,,
\end{equation}
with 
\beq\label{Utheta}
U(\Theta)= U_0(1-\Theta)^2\,, \hspace{5mm}
U_{\Lambda_0}(\Theta,\Phi)=\frac{{\Lambda_0}}{c^2} \Theta^2 e^{-\frac{4\gamma\Phi}{M_P}}\,,
\eeq
and 
\beqa
U_0\equiv\frac{\lambda \, a^2 M_P^4}{4}\,, \,\hspace{5mm} 
a \equiv \frac{1+6 \kappa}{\kappa}\,, \, \hspace{5mm} 
c\equiv\frac{\kappa}{\kappa_c}\gamma^2\,.
\eeqa
For sufficiently large values of $\Phi$, the symmetry-breaking potential $U_{\Lambda_0}$ 
becomes negligible as compared to $U$. In this limit, the field $\Phi$ is only derivatively 
coupled and the action acquires an emergent shift symmetry $\Phi\to\Phi+C$ with $C$ a constant. 
This symmetry can be understood as the non-linear realization of the original scale invariance and allows to interpret 
the field $\Phi$ as the Goldstone boson, or dilaton, associated to its spontaneous symmetry breaking. 
The dominant term $U$ approaches a constant value $U_0$ at small $\Theta$ values and allows for 
inflation with the usual chaotic initial conditions. 

Note that Eq.~\eqref{action_HD2} contains a non-canonical coefficient for the $\Theta$-field kinetic term,
\begin{equation}\label{poles} 
K(\Theta)=-\frac{M_P^2}{4\, \Theta}\left(\frac{1}{\kappa\Theta+c}+\frac{a}{1-\Theta}\right)\,,
\end{equation}
which could easily be removed by 
performing a simple field redefinition of the $\Theta$ field, $\bar\Theta\equiv \int \sqrt{K(\Theta)} d\Theta$. The 
structure of $K(\Theta)$ is, however, particularly enlightening. It contains three poles at $\Theta=0$, $\Theta=-c/\kappa$ and 
$\Theta=1$.  While the first two poles are potentially explored during inflation, the $\Theta=1$ pole is just 
a ``Minkowski'' pole in which the conformal factor \eqref{eq:conf} approaches one and the usual SM non-minimally coupled 
to gravity is approximately recovered. For the field values relevant for inflation, the 
``Minkowski'' pole can be safely neglected. In this limit, the field-derivative manifold in Eq.~\eqref{action_HD2} becomes 
a maximally symmetric space with Gaussian curvature $\kappa$ \cite{Karananas:2016kyt}.  As we will see below, this highly symmetric 
structure has a strong impact on the inflationary observables.

Given the above considerations a simple overall picture emerges. With the standard slow-roll 
initial conditions within the plateau region in Fig.~\ref{fig:potential}, the inflaton field $\Theta$ will tend to roll down the $U(\Theta)$ potential. 
While this happens ($\Theta\simeq 0$), the $\Phi$-field kinetic term is effectively suppressed and the dilaton ``freezes'' at its 
initial value $\Phi=\Phi_0$. This is an immediate consequence of scale-invariance. Indeed, as shown 
in Ref.~\cite{GarciaBellido:2011de}, the conservation of the Noether current associated to this continuous symmetry 
restricts the motion of the $h$ and $\chi$ fields to $\Phi={\rm constant}$ ellipsoidal trajectories in
the $\lbrace h,\chi\rbrace$ plane (cf.~Fig.~\ref{fig:potential}). 
This means that, in spite of dealing with a two-field model of inflation, 
no isocurvature perturbations nor non-Gaussianities are produced. For all 
practical purposes, the HD model behaves as a \textit{single-field} inflationary model \cite{GarciaBellido:2011de}. After
the end of inflation, entropy production takes place along the lines of Refs.~\cite{Bezrukov:2008ut,GarciaBellido:2008ab, Bezrukov:2011sz}.  
 Once the heating stage is complete, the field $\Theta$ settles down at the minimum of the $U$ potential. The expectation value 
 of $\Theta$ at this minimum, $\Theta_0=\Theta[h_0/\chi_0]$, translates into constant values for the masses of the SM
 particles produced during the heating stage. The radiation and matter dominated epochs proceed therefore 
 in the standard way. During these eras, the
almost massless scalar field $\Phi$ behaves essentially as a thawing quintessence field 
\cite{Wetterich:1987fm,Ratra:1987rm,Ferreira:1997hj,Caldwell:2005tm}.
In particular, it  stays ``frozen'' until the decreasing energy density in the SM sector equals 
the approximately constant term $U_{\Lambda_0}$. When that happens, the field starts rolling down this exponential
potential and the Universe enters a dark energy dominated era. In the following sections we discuss each of 
these epochs in detail.

\subsection{Inflation}

As argued in the previous section we can neglect the symmetry breaking potential $U_{\Lambda_0}$ and 
the ``Minkowski'' pole at $\Theta=1$ for the field values relevant for inflation. Given the 
resulting action, the inflationary observables can be computed using the standard techniques. Parametrizing the 
spectrum of scalar and tensor perturbations generated during inflation in the almost scale-invariant form \cite{Mukhanov:1990me}
\beq\label{defspec}
P_s=A_s\left(\frac{k}{k_*}\right)^{n_s-1+\frac12\alpha_s \ln (\frac{k}{k_*})}\,,\hspace{2mm} 
P_t=A_t\left(\frac{k}{k_*}\right)^{n_t},
\eeq
and using the slow-roll approximation for a non-canonical inflaton field $\Theta$, we get
\begin{eqnarray} 
 && \hspace{-5mm} A_s=\frac{1}{24\pi^2 M_P^4}\frac{U}{\epsilon}\,, \hspace{12mm} n_s=1+ 2\eta-6\epsilon\,,\label{Ansdef}\\ 
 && \hspace{-5mm} \alpha_s=8\epsilon(2\eta-3\epsilon)-2\delta^2, \hspace{5mm} r\equiv \frac{A_t}{A_s}=-8n_t=16\epsilon \label{asrdef}\,, 
\end{eqnarray}
with 
\beq\label{epsdef}
\epsilon\equiv\frac{M_P^2}{2 K}\left(\frac{U_{,\Theta}}{U}\right)^2\,, \hspace{10mm}
\eta \equiv \frac{M_P^2}{\sqrt{K}U}\left(\frac{U_{,\Theta}}{\sqrt{K}}\right)_{,\Theta} \,, 
\eeq
\beq
\delta^2 \equiv 
\frac{M_P^4 V_{,{\Theta}}}{ K V^2}\left[\frac{1}{\sqrt{K}}
\left(\frac{V_{,\Theta}}{\sqrt{K}}\right)_{,\Theta}\right]_{,\Theta}\,,
\eeq
the slow-roll parameters and $,\Theta$ denoting a derivative with respect to $\Theta$.
The inflationary observables \eqref{Ansdef} and \eqref{asrdef} 
are understood to be evaluated at a field value $\Theta_*\equiv \Theta(N_*)$. The quantity $N_*$ stands for the number 
of $e$-folds before the end of inflation at which the reference pivot scale $k_*$ in Eq.~\eqref{defspec} exited the 
horizon, $k_*=a_* H_*$. Using the first slow-roll parameter from Eq.~\eqref{epsdef} together with the standard
relation for the number of $e$-folds $N_*$ as a function of the non-canonical field $\Theta_*$, 
\beq
N_* =\frac{1}{M_P}\int_{\Theta_*}^{\Theta_{\rm E}} \frac{\sqrt{K}d \Theta}{\sqrt{2\epsilon}}\,,
\eeq
we obtain 
\beq\label{Nefolds}
N_*=\frac{1}{8c}\ln\left[\frac{\Theta_*}{\Theta_{\rm E}}\left(\frac{\kappa \Theta_{\rm E}+c}{\kappa\Theta_*+c}
  \right)^{1+\frac{c}{\kappa}}\right]\,,
\eeq
with
\begin{equation}
\Theta_{\rm E}=\frac{1-4c-2\sqrt{4c^2-2c-2\kappa}}{1+8\kappa}
\end{equation}
the value of the field at the end of inflation, $\epsilon(\Theta_{\rm E})\equiv1$. To invert Eq.~\eqref{Nefolds} 
we assume the ratio $c/\vert \kappa\vert$ to be small and perform a Taylor expansion 
around  $c/\vert \kappa\vert \approx 0$. In terms of the original fields $h$ and $\chi$, this corresponds to an inflationary 
dynamics which is essentially dominated by the Higgs component, i.e. with $\xi_h\gg \xi_\chi $ or equivalently with 
$\vert \kappa\vert\simeq \vert\kappa_c\vert$, cf. Eq.~\eqref{kappadef}. To the lowest order in $c/\vert\kappa\vert$ we obtain the 
following analytical expressions for
the amplitude of the primordial spectrum of scalar perturbations, 
\begin{equation}\label{As}
A_s
 =\frac{\lambda \sinh^{2} \left(4 c N_* \right)}{1152\pi^2 \xi^2_{\rm eff}\,  c^2} \,,\hspace{10mm} 
 \xi_{\rm eff}\equiv\frac{1}{\sqrt{6 a^2 \vert\kappa_c\vert}}\,,
\end{equation}
the spectral tilt and its running
\begin{eqnarray}
n_s &=& 1-8\,c \coth\left(4 c N_*\right)\,,  \label{nsr1}  \\ 
\alpha_s &=& - 32\, c^2 \csch^{2} \left(4 c N_* \right)\,, \label{nsr1b} 
\end{eqnarray}
and the tensor-to-scalar ratio
\beq\label{nsr2}
r=  \frac{32\, c^2}{\vert \kappa_c\vert} \csch^{2} \left(4 c N_*\right)\,.
\eeq 
Note that the spectral tilt is bounded from above and that the observables \eqref{nsr1}-\eqref{nsr2} are non-trivially related,
\beq
\label{eq:bound2}
n_s\lesssim 1-\frac{2}{N_*}\,,\hspace{5mm} r\leq \frac{(n_s-1)^2}{2\vert\kappa_c\vert}\,,
\hspace{5mm}\alpha_s=-\vert\kappa_c\vert \, r\,.
\eeq 
As long as the quantity $4 c N_*$ is smaller than one, the 
series expansions of the hyperbolic functions in Eqs.~\eqref{nsr1}-\eqref{nsr2} rapidly converge and 
we can further approximate them by 
\beqa\label{eq:bound}
n_s \simeq 1-\frac{2}{N_*} \,,  \hspace{5mm} 
\alpha_s \simeq  - \frac{2}{N_*^2}\,, \hspace{5mm} r\simeq  
\frac{2}{\vert \kappa_c\vert N_*^2} \label{nsr2app}\,.
\eeqa 
The structure of these equations is a natural 
consequence of the pole structure in Eq.~\eqref{poles} \cite{Karananas:2016kyt}. In the 
limit $c\to 0$, the pole in this expression becomes approximately quadratic. This behavior translates into an
exponential flattening of the $U$ potential when written in terms of 
a canonically normalized field $\bar\Theta\equiv \int \sqrt{K(\Theta)} d\Theta$. The quadratic pole 
acts as an attractor driving the inflationary observables towards the values 
\eqref{nsr2app}.  Note that for not too small $\vert\kappa_c\vert$, the tensor-to-scalar ratio is highly suppressed, 
 $r\sim {\cal O}(10^{-3})$. This shares similarities with the $\alpha$-attractors discussed in  
Refs.~\cite{Ferrara:2013rsa,Kallosh:2013yoa,Galante:2014ifa} (for a connection between this approach 
and the existence of stationary points along the inflationary potential see Ref.~\cite{Artymowski:2016pjz}). 

For $4cN_*$ larger than one the inflationary observables \eqref{nsr1}-\eqref{nsr2} approach the 
\textit{asymptotic} values 
\beqa
n_s \simeq  1-8c \,,  \hspace{10mm} \alpha_s \simeq  0\,, \hspace{10mm} r\simeq 0 \label{nsr2app2}\,.
\eeqa 
This limit is again a natural consequence of the pole structure in Eq.~\eqref{poles} ~\cite{Karananas:2016kyt}. Indeed, 
for $c\neq 0$, the inflationary pole at $\Theta=0$ is no longer reachable and we are left with a linear 
pole structure. As shown in Ref.~\cite{Terada:2016nqg}, the linear pole 
also acts as an effective attractor driving the inflationary observables towards the values \eqref{nsr2app2}.  

\subsection{Heating and entropy production}\label{sec:heating}
 The precise value of $N_*$ in Eqs.~\eqref{As}-\eqref{nsr2} depends on the duration of the heating stage. 
The heating stage in Higgs-Dilaton inflation coincides with that in Higgs inflation \cite{Bezrukov:2008ut,GarciaBellido:2008ab}, 
up to a negligible dilaton production\footnote{The Goldstone boson nature of the dilaton makes it a potential 
candidate for contributing to the effective number of relativistic degrees of freedom at Big Bang Nucleosynthesis 
and recombination. However, the extremely small production of this component during the reheating stage translates 
into a number of degrees of freedom  very close to the Standard Model one \cite{GarciaBellido:2012zu}.  For all practical purposes, the 
dilaton \textit{excitations} remain elusive and completely undetectable by any particle physics experiment 
or cosmological observation. The only remnant of the dilaton \textit{field} is
a dynamical dark energy stage, cf. Section \ref{sec:DE}.} associated to the non-canonical kinetic term in Eq.~\eqref{action_HD2} 
\cite{GarciaBellido:2011de,GarciaBellido:2012zu}. After the end of inflation 
the $\Theta$ field starts to oscillate around the minimum of $U$. These oscillations lead to the 
production of $W^{\pm}$ and $Z$ gauge bosons which tend to decay upon production into the Standard Model quarks and leptons.
The decay probability is proportional to the gauge boson effective mass, which depends itself on the $\Theta$-field 
expectation value. The large amplitude of the dynamical field $\Theta$ at the onset of the heating stage translates 
into a very efficient decay which tends to deplete the gauge boson occupation numbers within a single oscillation
and delays the onset of parametric resonance \cite{Bezrukov:2008ut,GarciaBellido:2008ab}.  As the Universe expands, the 
average value of $\Theta$ decreases and eventually becomes small enough as to allow for the gauge bosons to accumulate.  When 
that happens, the system enters into a parametric resonance regime and eventually backreacts into the inflaton condensate.  From there on until thermalization, 
the evolution of the system is very  non-linear and non-perturbative and one must rely on lattice 
simulations~\cite{Repond:2016sol}. The different analytical and numerical considerations in 
Refs.~\cite{Bezrukov:2008ut,GarciaBellido:2008ab,GarciaBellido:2011de,Repond:2016sol} seem to indicate 
that heating in Higgs-Dilaton inflation takes place rather fast, leading to a relatively well-constrained number 
of $e$-folds,\footnote{We use the Planck satellite pivot scale $k_*=0.05/$Mpc rather than $k_*=0.002/$Mpc. This 
translates into a shift of roughly three $e$-folds  with respect to the estimates in Ref.~\cite{GarciaBellido:2011de}.}
$60 \lesssim N_*\lesssim 62\,.$

\subsection{Dark energy}\label{sec:DE}

 Given the value of the Higgs self-coupling at the inflationary 
 scale\footnote{For details cf.~Ref.~\cite{Bezrukov:2014ipa}. For 
 the HD model the behavior near the critical point has been studied in Ref.~\cite{Rubio:2014wta}.} and the aforementioned 
 duration of the heating stage, the values of $\vert \kappa\vert\simeq \vert \kappa_c\vert$ and $c$ (or equivalently of the non-minimal couplings $\xi_h$ and $\xi_\chi$) can be fixed by comparing 
 the inflationary predictions \eqref{As}-\eqref{nsr2} with cosmic microwave background (CMB) data. The free parameters of the model then become 
 completely determined. This allows us to make specific predictions for any subsequent period in the 
 evolution of the Universe. In particular, it is possible to derive an extremely appealing connection between
 inflation and the present dark energy dominated era. Establishing this connection is the purpose of this section.

After the end of the reheating stage, the field $\Theta$ will relax towards the minimum of $U$. When this
happens, the Higgs-Dilaton action~\eqref{action_HD2} boils down to the simple Lagrangian density
\begin{eqnarray}\label{action_HDde}
\frac{\cal L}{\sqrt{-g}}&\simeq &  \frac{M_P^2}{2}R 
-\frac{1}{2}(\partial \Phi)^2 -\frac{{\Lambda_0}}{c^2} e^{-\frac{4\gamma\Phi}{M_P}}\,,
\end{eqnarray}
supplemented by the Lagrangian density for the matter and radiation components produced during 
the heating stage ~\cite{Bezrukov:2008ut,GarciaBellido:2008ab,Repond:2016sol}. 

If ${\Lambda_0}>0$, the potential for the dilaton field is of the runaway-type and can 
support an accelerated expansion of the Universe while driving $\Phi\rightarrow\infty$. 
This type of potential was first considered in the seminal papers \cite{Wetterich:1987fk,Wetterich:1987fm}. 
Regarding the late time acceleration of the Universe, the \textit{cosmon scenario} 
presented there displays some formal similarities with
the HD model but at the same time some conceptual differences. In Refs.~\cite{Wetterich:1987fk,Wetterich:1987fm} the runaway potential 
in Eq.~\eqref{action_HDde} appears as a consequence of the dilatation anomaly and therefore of the 
explicit breaking of scale invariance. In the HD model, scale-invariance is assumed to be an 
exact symmetry at the quantum level and the exponential potential appears as a consequence of the unimodular 
 character of gravity. 

The quintessence potential in the Higgs-Dilaton scenario is nearly flat and the quintessence fluid 
 started to dominate only recently. We are thus dealing with a thawing quintessence 
 scenario~\cite{Caldwell:2005tm,Ferreira:1997hj}. 
The evolution equations for the dilaton/matter system can be written as 
\cite{Wetterich:2003qb,Scherrer:2007pu,Chiba:2012cb} 
\begin{eqnarray}
&& \hspace{-10mm}\frac{w'}{1-w}=-3(1+w)+4\gamma\sqrt{3(1+w)\Omega_{\rm DE}}\label{weq}\,,  \\
&& \hspace{-10mm}\,\Omega'_{\rm DE}=-3\Omega_{\rm DE} (1-\Omega_{\rm DE})w \label{Oeq}\,,
\end{eqnarray}
with $\Omega_{\rm DE}$ the dark-energy energy density parameter associated to the dilaton 
field $\Phi$, $w$ its effective equation-of-state parameter and the primes denoting derivatives with respect
to the number of $e$-folds  $N=\ln a$.  During matter and radiation domination the density parameter $\Omega_{\rm DE}$ 
is small and the last term in  Eq.~\eqref{weq} can be safely neglected. The dilatonic dark energy component then behaves as a (subdominant) 
 cosmological constant with an equation-of-state parameter $w\simeq -1$.  However, since the dilaton energy
 density remains approximately constant, the fraction $\Omega_{\rm DE}$ will eventually become relevant. 
As shown in Refs.~\cite{Wetterich:1987fm,Copeland:1997et,Ferreira:1997hj}, the set of equations \eqref{weq} and \eqref{Oeq} admits a 
stable fixed point, 
\beq 
\Omega_{\rm DE}=1\,,\hspace{10mm} 1+w=\frac{16\gamma^2}{3}\,,
\eeq
which leads to acceleration if $\gamma<1/(2\sqrt{2})$. Indeed, in the approximation $1 + w \ll 1$, we can easily integrate 
Eqs.~\eqref{weq} and \eqref{Oeq} to obtain \cite{Scherrer:2007pu}
\beq \label{eqn:eos_omega_quintessence}
1+w= \frac{16 \gamma^2}{3} F(\Omega_{\rm DE})\,,  \hspace{5mm}
\Omega_{\rm DE}=\frac{1}{1+\Delta_{0}\,a^{-3}}\,,
\eeq
with
\beq 
F(\Omega_{\rm DE})=\left[\frac{1}{\sqrt{\Omega_{\rm DE}}}-\Delta\tanh^{-1} \sqrt{\Omega_{\rm DE}}\right]^2\,,
\eeq 
and 
\beq
\Delta  \equiv \frac{1-\Omega_{\rm DE}}{\Omega_{\rm DE}}\,,\hspace{10mm} 
\Delta_0  \equiv \frac{1-\Omega_{\rm DE,0}}{\Omega_{\rm DE,0}}\,.
\eeq
The subscript $0$ marks quantities evaluated today.
The function $F(\Omega_{\rm DE})$ is a monotonically increasing function smoothly interpolating between $F(0)=0$ in 
the deep radiation and matter dominated eras and $F(1)=1$ in the asymptotic dark energy dominated era. The present value
of the dark-energy equation-of-state parameter follows directly from Eq.~\eqref{eqn:eos_omega_quintessence},
\beq
  \label{eqn:eos_omega_quintessence2}
\frac{1+w_0}{1+w(a)}= \frac{F(\Omega_{\rm DE,0})}{F(\Omega_{\rm DE})}.
\eeq

\subsection{Consistency relations}\label{sec:consistency_rel}
Note that the equation-of-state parameter in Eq.~\eqref{eqn:eos_omega_quintessence} depends only on $\gamma$, which 
coincides with $\sqrt{c}$ in the $\vert\kappa\vert\approx \vert\kappa_c\vert$ approximation. 
Combining this expression with Eq.~\eqref{nsr1} we get the following consistency relation between the spectral 
tilt of scalar perturbations generated during inflation and the equation-of-state parameter of dark energy, namely
\beq\label{nswcons}
n_s = 1- \frac{2}{N_*}X \coth X\,,
\eeq
with
\beq
X\equiv 4cN_*=\frac{3 N_*(1+w)}{4 F(\Omega_\textrm{DE})}\,.
\eeq
A similar consistency relation can be derived for the running of the spectral tilt, 
$\alpha_s=-\vert \kappa_c\vert r$, and  the tensor-to-scalar ratio 
\beq\label{rascons}
r=\frac{2}{\vert\kappa_c \vert N_*^2}  X^2\sinh^{-2} X\,.
\eeq 
The validity of the consistency relations \eqref{nswcons} and \eqref{rascons} is restricted to the range of validity of the 
approximation $c/\vert \kappa\vert \ll 1$. The derivation presented in this paper allows for a 
straightforward generalization to general scale-invariant scenarios as those considered 
in Ref.~\cite{Karananas:2016kyt}. For the particular model under consideration, the constant 
$\vert\kappa\vert\simeq\vert\kappa_c\vert$ is determined by the amplitude of the primordial power spectrum \eqref{As} once the 
value of the Higgs self-coupling at the inflationary scale is specified. For the values of $\lambda$ following from the 
usual SM renormalization group running \cite{Bezrukov:2012sa,Degrassi:2012ry,Buttazzo:2013uya,Bezrukov:2014ina}, one gets 
$\xi_{\rm eff}\simeq \xi_h\gg 1$, which leads to a value $\vert\kappa\vert\simeq \vert\kappa_c\vert\simeq 1/6$ for 
the field-derivative space curvature. In this limit, the 
expressions \eqref{nswcons}  and \eqref{rascons} reduce to those in Ref.~\cite{GarciaBellido:2011de}.

Whether the consistency relations \eqref{nswcons} and \eqref{rascons} remain unaltered or rather become modified in 
the presence of quantum corrections depends on the particular UV completion of the Standard Model non-minimally coupled to 
gravity. A potential UV completion respecting the symmetries of the non-renormalizable action \eqref{eqn:hdm_lagrangian} 
was conjectured in Refs.~\cite{Shaposhnikov:2008xb,Shaposhnikov:2008xi}. The bottom-up approach of 
Refs.~\cite{Armillis:2013wya,Gretsch:2013ooa} shows that it is indeed possible to remove all the divergencies in the 
theory while keeping scale invariance intact. The price to pay is the lack of renormalizability \cite{Shaposhnikov:2009nk}, which 
does not seem to be a strong requirement given that gravity itself is non-renormalizable. For scale-invariant UV completions, 
the consistency relations \eqref{nswcons} and \eqref{rascons} remain unaltered \cite{Bezrukov:2012hx}. Deviations from these consistency 
relations are only expected in the so-called critical regime in which the Higgs-Dilaton potential develops an inflection 
point along the inflationary trajectory \cite{Rubio:2014wta} (see also Refs.~\cite{Bezrukov:2014bra,Hamada:2014iga,Bezrukov:2017dyv}).

\section{Current data}\label{sec:data1}

\begin{table*}
	\centering{}
	\begin{ruledtabular}
		\begin{tabular}{llrccc}
		\textbf{Parameter}  & Description & \textbf{Prior range} & \textbf{$\Lambda$CDM} & \textbf{HD}  & \textbf{$w$CDM}\tabularnewline
		\hline 
		$\omega_b=\Omega_b h^2$ & Baryon density today & $[0.005, 0.1]$  & \textbullet & \textbullet & \textbullet \tabularnewline
		$\omega_{cdm}=\Omega_{cdm} h^2$ & CDM density today & $[0.001, 0.99]$  & \textbullet & \textbullet & \textbullet \tabularnewline
		$h$  & Dimensionless Hubble constant & $[0.1, 2]$ & \textbullet & \textbullet & \textbullet \tabularnewline
		$\ln\left(10^{10}A_s\right)$ & Amplitude of primordial curvature perturbations & $[2, 4]$  & \textbullet & \textbullet & \textbullet \tabularnewline
		$\tau_{reio}$ & Optical depth due to reionization & $[0.01, 0.8]$  & \textbullet & \textbullet & \textbullet \tabularnewline
		$n_s$  & Spectral tilt &$[0.8, 1.2]$  & \textbullet &  & \textbullet \tabularnewline
		$r$ & Tensor-to-scalar ratio & $[0.0, 0.5]$  & \textbullet &  & \textbullet \tabularnewline
		$w_0$ & Current value of the dark energy equation of state & $[-1, 0]$  &  & \textbullet & \textbullet \tabularnewline
		$N_*$ & Number of $e$-folds of inflation w.r.t. $k=0.05 \text{ Mpc}^{-1}$ & $(60\pm2.5)$ & & \textbullet & \tabularnewline
		$\Sigma m_\nu$ & Sum of neutrino masses & $[0.0, 5.0]$  & \textbullet & \textbullet & \textbullet \tabularnewline
		\end{tabular}
	\end{ruledtabular}
	\caption{\label{tab:priors} The priors used in the MCMC analysis. The dots indicate the parameters that are 
	varied as independent parameters in the given model. For the number of $e$-folds $N_*$, we implement 
	a strong Gaussian prior (see the text). We choose flat priors for the other parameters.}
\end{table*}

The Higgs-Dilaton consistency relations derived in the previous section bring us to an unusual situation in which some 
cosmological observables that are customarily understood as independent parameters become related in a rather 
non-trivial manner. In a Higgs-Dilaton cosmology, a dynamical dark energy component is inevitably 
associated to deviations from the $\alpha$-attractor predictions in Eq.~\eqref{eq:bound}. To understand the impact of the 
 consistency relations on cosmological observables we will consider three different scenarios:
\begin{enumerate}
 \item A standard $\Lambda$CDM cosmology. 
 \item A HD model satisfying the consistency relations \eqref{nswcons}-\eqref{rascons} between the dark-energy 
 equation-of-state parameter, the spectral tilt and the tensor-to-scalar ratio.
\item A $w$CDM model satisfying the thawing quintessence condition \eqref{eqn:eos_omega_quintessence2} 
but without any additional constraints on the primordial power spectra.
\end{enumerate}
By comparing the HD model to $\Lambda$CDM, we will explore how the observables in the former scenario differ 
from those in the standard cosmological picture. The comparison of the HD model with a $w$CDM cosmology 
will allow us to isolate the impact of the consistency relations.

\subsection{Parameters and priors}

For each of the aforementioned models we follow the conventional Bayesian approach for the estimation of cosmological
parameters and perform a Markov Chain Monte Carlo (MCMC) analysis. The iterative  MCMC algorithm samples the 
parameter space weighted by a prior distribution by constructing a Markov chain whose equilibrium distribution
is the target posterior distribution. The quality of the sampling improves with the number of steps. After sampling
for a sufficiently long time, the chain reaches equilibrium and the samples can be regarded as those from the desired
posterior distribution. The varying parameters  in the $\Lambda$CDM, HD and $w$CDM scenarios are summarized in Table 
\ref{tab:priors}. For all cases, we assume an exactly flat Universe with $\Omega_{\rm K}=0$, three degenerate massive neutrinos 
and no additional sterile neutrino species. In the HD case we vary only the present dark-energy equation-of-state parameter and 
use  the consistency relations \eqref{nswcons}-\eqref{rascons} to compute the spectral tilt and the tensor-to-scalar 
ratio from it. Doing so implies
adding the number of $e$-folds of inflation as a free parameter. For the $w$CDM case we adopt the dark-energy 
evolution \eqref{eqn:eos_omega_quintessence2} but without implementing any additional constraint or consistency relation on the 
primordial power spectra. Since the spectral tilt and the tensor-to-scalar ratio are free parameters in this case, the 
number of $e$-folds does no longer play any role.

As argued in Section \ref{sec:heating}, all HD model estimates predict a rapid entropy production after inflation leading to a number 
of $e$-folds $N_*\simeq 60$, with an uncertainty of a few $e$-folds. To account for the theoretical knowledge of this heating stage 
we implement a strong theoretical prior for the number of $e$-folds of inflation. In particular, we choose a Gaussian prior with central value $\mu=60$ and standard deviation  
$\sigma=2.5$. For the rest of the parameters we implement flat priors with boundaries as given in Table~\ref{tab:priors}.
 Since the HD model does not allow for a phantom behavior $w<-1$, we restrict the current dark-energy equation-of-state 
 to be larger than $-1$.  We also apply this restriction to the $w$CDM case to 
 compare the two models on equal footing.

\subsection{MCMC analysis}\label{sec:dataMCMCanalysis}
To perform the MCMC analysis we make use of the MontePython code \cite{audren:montepython}. 
This code is a MCMC cosmological parameter extraction code containing the most recent observational likelihoods and
interfaced with the Cosmic Linear Anisotropy Solver Software (CLASS) for the computation of
cosmological observables \cite{lesgourgues:class}. To implement the constraints described in Section 
\ref{sec:consistency_rel} we have modified the CLASS code in two ways. First,
we have implemented the thawing quintessence evolution  \eqref{eqn:eos_omega_quintessence2} for 
the equation-of-state parameter $w(a;w_0, \Omega_{\rm DE,0})$. This evolution is used both in 
the HD scenario and in the $w$CDM case. Second, we have modified the initial perturbations in the HD case to choose 
the spectral tilt and the tensor-to-scalar ratio as a function of the equation-of-state parameter $w_0$ and the number of $e$-folds $N_*$, 
in accordance with the consistency relations \eqref{nswcons}-\eqref{rascons}. 

We consider the following data sets: 
i) the 2015 Planck high-multipole TT likelihood, the Planck low-multipole polarization and temperature
likelihood as well as the Planck lensing likelihood \cite{Aghanim:2015xee}, 
ii) the Keck/Bicep2 likelihood data release 2015 \cite{bicep2}, 
iii) the Joint Lightcurve Analysis \cite{betoule:jla} and  
iv) baryonic acoustic oscillations data sets from 6dF, BOSS and SDSS \cite{Beutler:2011hx, anderson:boss, Ross:2014qpa}. 
From these likelihoods we get 26 additional nuisance parameters.

Our MCMC analysis is similar in spirit to the one performed in Ref.~\cite{Trashorras:2016azl}. However
we differ from this work in multiple ways. First, we present an additional comparison of the HD model to $\Lambda$CDM. Second, by 
using the parametrization \eqref{eqn:eos_omega_quintessence2} for $w(a)$ we capture the \textit{full} 
thawing quintessence behavior  of the Higgs-Dilaton model and avoid introducing an 
additional parameter $w_a$ to account for the temporal dependence of the dark energy equation-of-state. Third, 
by constraining the number of $e$-folds, we only allow for values of this quantity that are in agreement 
with theoretical expectations. Finally, we do \textit{not} allow for a phantom behavior with $w<-1$. 
A violation of the null energy condition in the HD model would imply a negative value for the non-minimal coupling $\xi_\chi$ 
[cf. Eq.~\eqref{eqn:eos_omega_quintessence}] and therefore a negative Planck 
mass square, $M_P^2\simeq \xi_\chi \chi_0^2$ once the Higgs field settles down at 
the electroweak vacuum $h_0^2\simeq \alpha\chi_0^2$, which is obviously unacceptable. 

\begin{figure*}
 \includegraphics[width=\textwidth]{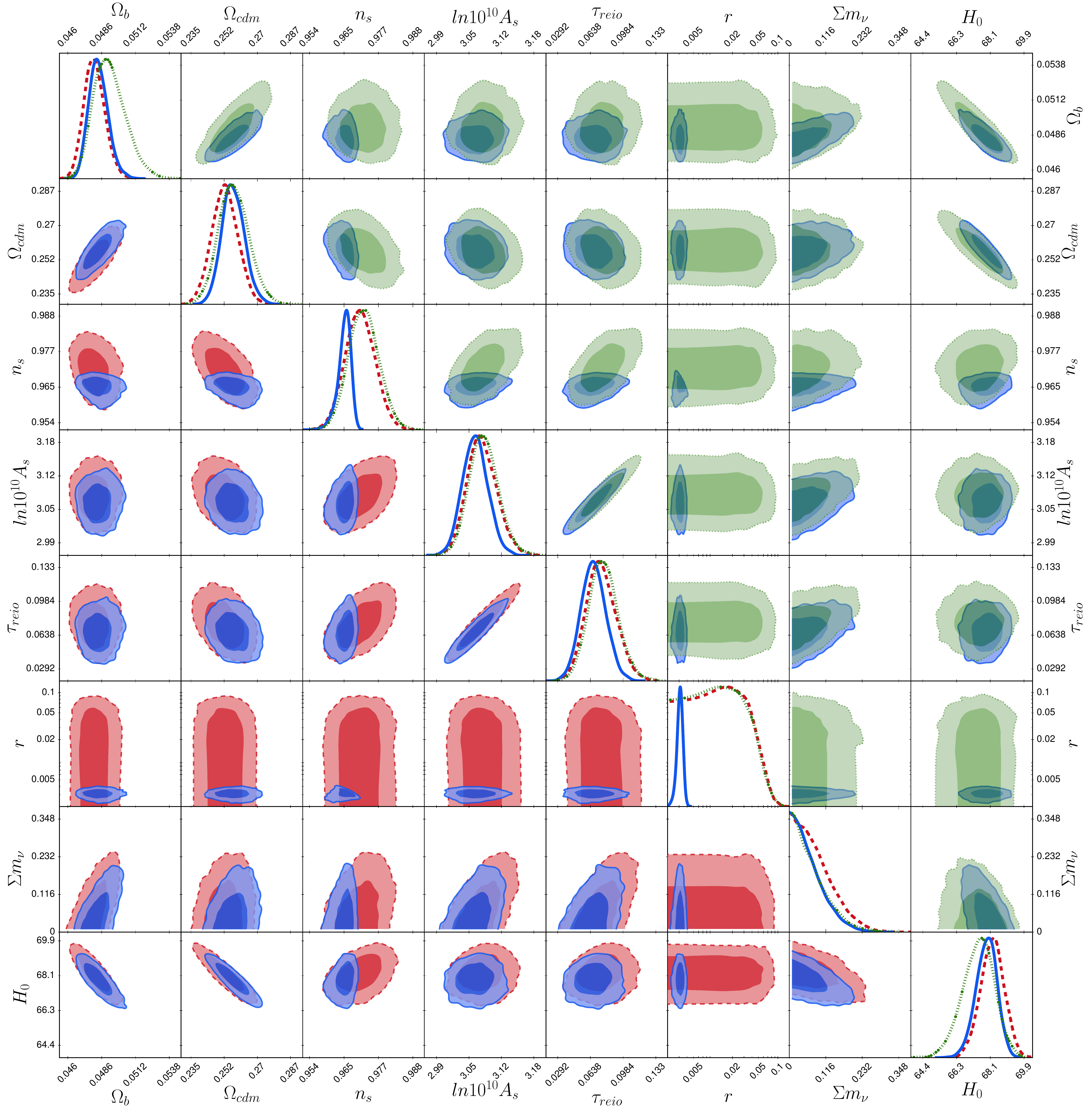}
 \caption{\label{fig:mcmc} Results of the MCMC run for different scenarios. The plots in the bottom left 
 corner compare $\Lambda$CDM (red, dashed) with the HD model (blue, solid). The comparison of the latest scenario with $w$CDM 
 (green, dotted) is shown in the upper right corner. Note that the tensor-to-scalar ratio $r$ is plotted on a 
 logarithmic scale (see Fig.~\ref{fig:mcmc} for a detailed view of the $r$-$n_s$ contours in the HD model) .}
\end{figure*}

\begin{figure*}
  \includegraphics[scale=0.24]{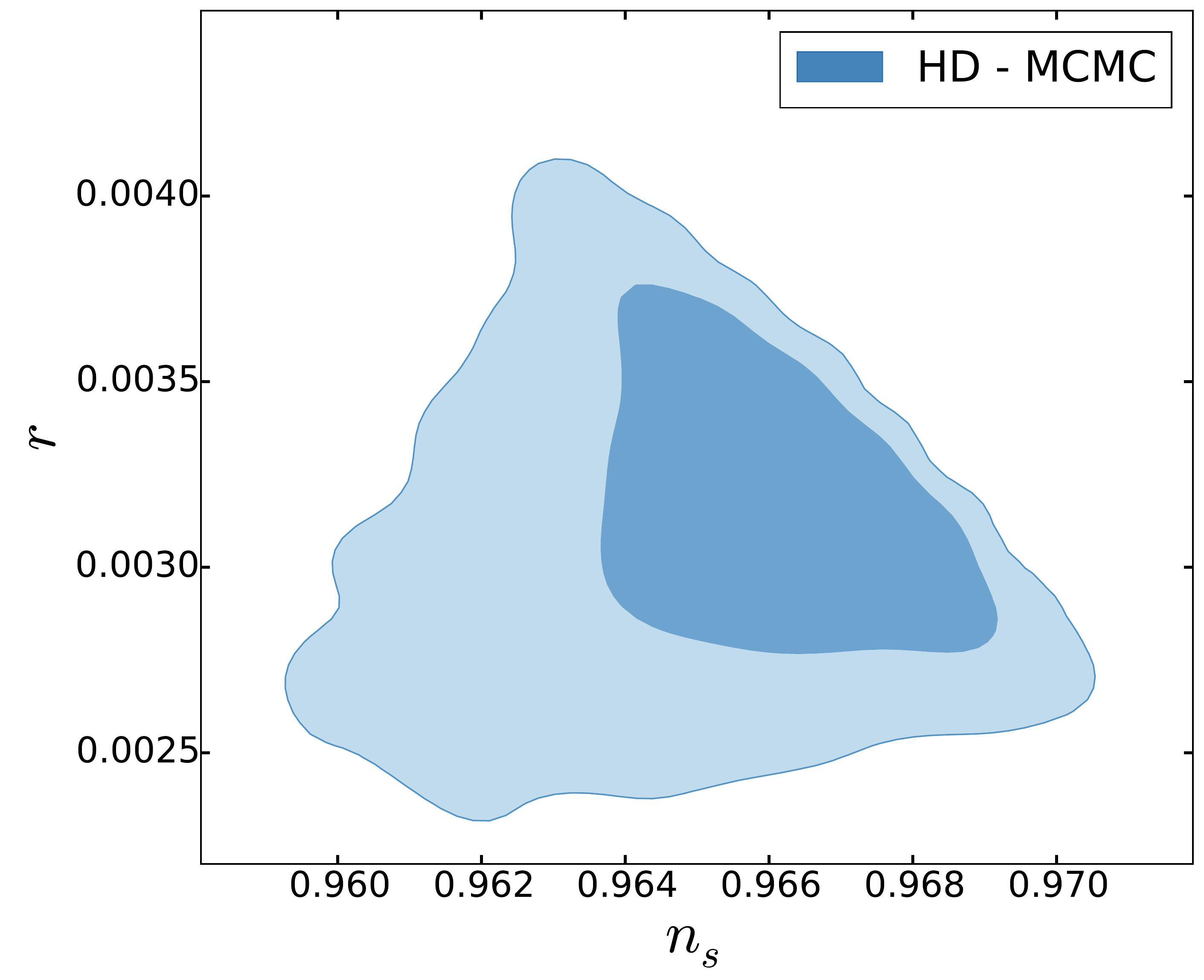} \hspace{15mm}
\includegraphics[scale=0.25]{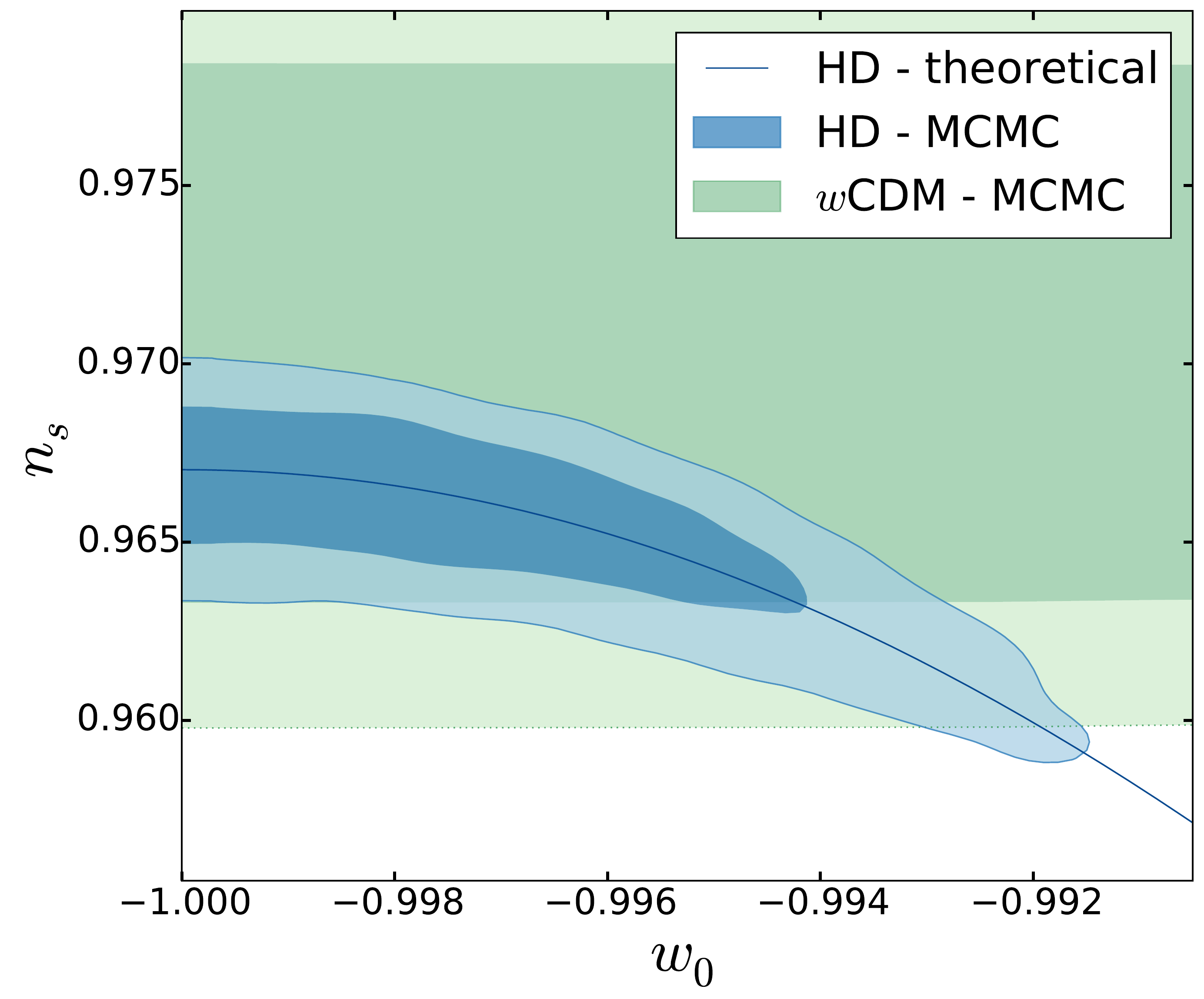}
 \caption{\label{fig:consplots}
(Left) Detailed view of the $1\sigma$ and $2\sigma$ contours for the spectral tilt and 
tensor-to-scalar ratio in a HD cosmology. These values are obtained by means of the consistency relations 
\eqref{nswcons}-\eqref{rascons}. This $r$-$n_s$ consistency relation strongly restricts the possible values for the spectral 
tilt $n_s$ and the tensor-to-scalar ratio $r$ as compared to the $\Lambda$CDM and $w$CDM scenarios 
(cf. Fig.~\ref{fig:mcmc}). 
(Right) Detailed view of $1\sigma$ and $2\sigma$ contours for the spectral tilt and the present dark-energy 
equation-of-state in the HD model (blue, solid) and the $w$CDM scenario (green, dotted). The blue solid line stands 
for the expected HD consistency relation \eqref{nswcons} evaluated at the mean values of the MCMC run 
($N_*=60.7 $, $\Omega_{\rm DE,0}=0.696$).}
\end{figure*}

\subsection{MCMC results}\label{sec:MCMCresults}
In total we ran eight chains per model to obtain approximately $\sim 2\times10^{5}$ 
samples per model. 
The results are depicted in Fig.~\ref{fig:mcmc} and \ref{fig:consplots}, where we present histograms and correlation contours 
for the relevant cosmological parameters in the $\Lambda$CDM (red, dashed), HD (blue, solid) and $w$CDM (green, dotted) scenarios. To facilitate the comparison between the 
different models we converted the free parameter $w_0$ in the HD case  
to the derived parameters $n_s$ and $r$ by means of the consistency relations \eqref{nswcons}-\eqref{rascons}.

Qualitatively one finds that the HD model resembles a \lcdm{} cosmology with additional 
constraints on the spectral tilt and the tensor-to-scalar ratio, cf. Eq.~\eqref{eq:bound2}. 
At current precision, the present dark energy equation-of-state parameter in the HD model is still compatible with 
a cosmological constant ($w_0=-1$). Note, however, that our best fit value $w_0=-0.997\pm 0.003$ differs from that in 
Ref.~\cite{Trashorras:2016azl} due to the physical requirement of having a well-defined graviton propagator at all 
field values. In particular, our dark-energy equation-of-state parameter is \textit{not} phantom. 
The comparison of the HD model with a $w$CDM cosmology reveals also that the baryon fraction 
$\Omega_b$ and the Hubble constant $H_0$ are better constrained in the presence 
of the consistency relations  \eqref{nswcons}-\eqref{rascons}, with ranges comparable to those in the standard 
$\Lambda$CDM case.

\begin{figure}
  \includegraphics[scale=0.24]{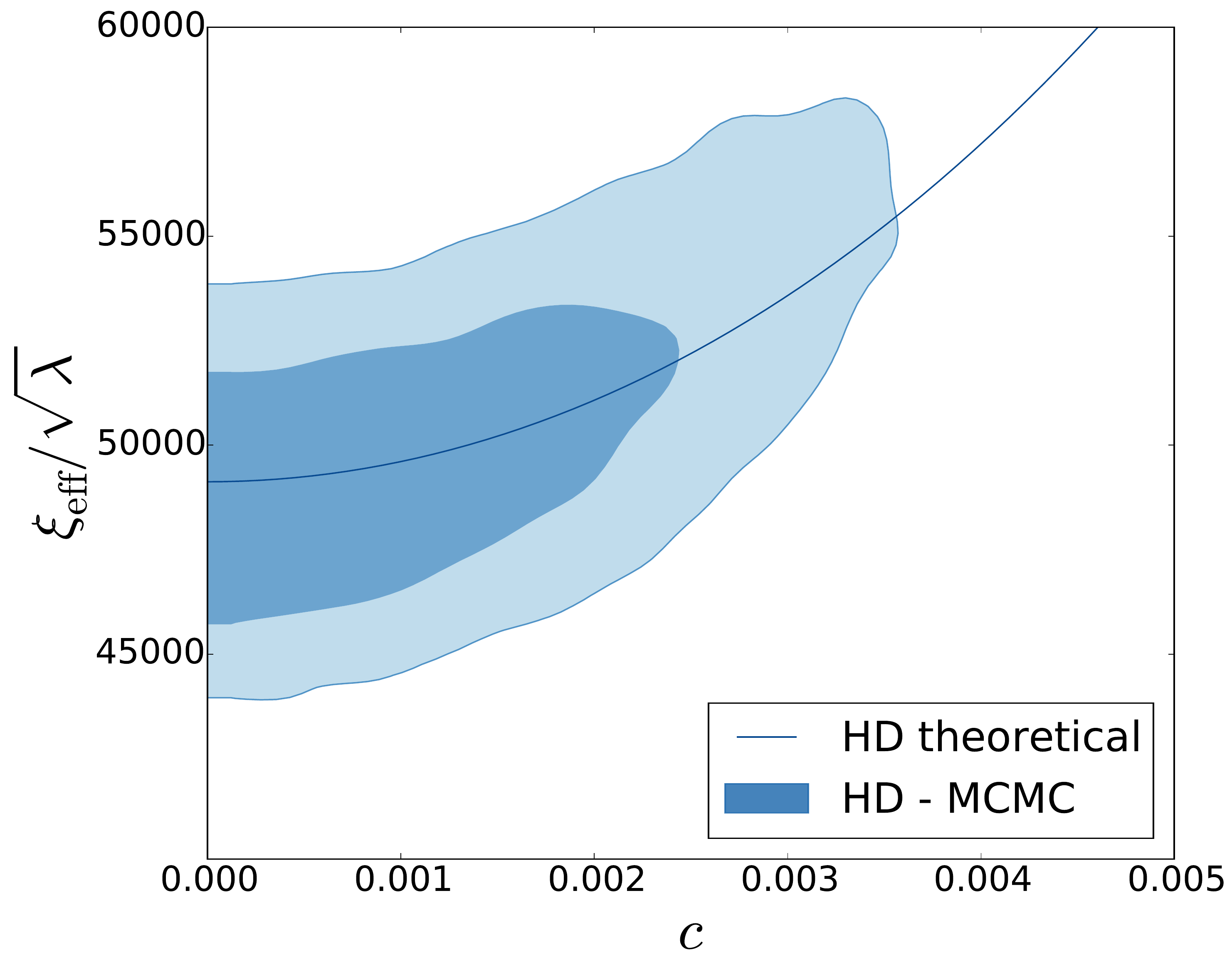} 
 \caption{\label{fig:paramcontours}
1$\sigma$ and 2$\sigma$ constraints on the model parameters $\xi_{\rm eff}$ and $c$. The contours are obtained by combining  
Eqs.~\eqref{As} and \eqref{nsr1} with the observationally allowed regions for the curvature power spectrum and the spectral 
tilt, cf.~Fig.~\ref{fig:mcmc}. 
The blue solid line stands for the expected HD relation evaluated at the mean values of the MCMC run ($N_*=60.7 $, $\ln(10^{10}A_s) = 3.07$).
Note that for not to small values of the Higgs self-coupling, e.g. $\lambda\sim {\cal{O}}(10^{-3})$, the condition $\xi_{\rm eff}/\sqrt{\lambda}\gg1$ translates into a large 
value of the Higgs non-minimal coupling, $\xi_{\rm eff}\simeq \xi_h\gg 1$.}
\end{figure}

Given the contours for the amplitude of the curvature power spectrum $A_s$ and the spectral tilt $n_s$ we can
immediately derive constraints for the model parameters $\xi_{\rm eff}$ and $c$ using Eqs.~\eqref{As} and \eqref{nsr1}. The result is shown in 
Fig.~\ref{fig:paramcontours}. The allowed values for the amplitude 
give rise to a band shape in the $\lbrace\xi_{\rm eff}/\sqrt{\lambda},c\rbrace$ plane, whose upper limit is determined 
by the minimal values of the spectral tilt compatible with its $1\sigma$ and $2\sigma$ contours. As anticipated in section
\ref{sec:consistency_rel}, a value of the Higgs-self coupling $\lambda$ compatible with the usual SM renormalization 
group running, $\lambda\sim {\cal{O}}(10^{-3})$, translates into a sizable value of the effective coupling $\xi_{\rm eff}$. 
In that limit, $\xi_{\rm eff} \simeq \xi_h\gg 1$ and $\vert \kappa_c\vert\simeq 1/6$.

To perform a more quantitative comparison among the different models we compute the Bayes factor for each scenario. 
Assuming that all models are \textit{a priori} equally probable, $\pi(M_{\Lambda \text{CDM}})/\pi(M) = 1$, 
the Bayes factor 
\beq
  B(M) = \frac{p(\mathbf{x}|M)}{p(\mathbf{x}|M_{\Lambda \text{CDM}})} = \frac{\pi(M_{\Lambda \text{CDM}})}{\pi(M)} 
  \frac{p(M|\mathbf{x})}{p(M_{\Lambda \text{CDM}}|\mathbf{x})}
\eeq
measures the probability $p(M|\mathbf{x})$ of a model $M$ given current data $\mathbf{x}$ as compared to the probability 
$p(M_{\Lambda \text{CDM}}|\mathbf{x})$ of a $\Lambda$CDM scenario given the same data set. 
In interpreting this quantity we adopt the Kass and Raftery scale \cite{kass1995bayes}, a revision of Jeffreys scale. 
In this scale a value of $|\Delta\ln B| > 3$ is understood as strong statistical evidence, 
translating into a relative probability of approximately $1/20$ between the two models under consideration.
To estimate the evidence $p(M|\mathbf{x})$ directly from our MCMC chains we use the nearest neighbor 
approximation\footnote{The method assumes that individual points in the chain are 
independent. For a chain from a MCMC run this is clearly not the case. 
We checked that our qualitative results do not depend on the amount of thinning applied to the chains.}
introduced in Refs.~\cite{Heavens:2017afc, Heavens:2017hkr} and marginalize over the nuisance parameters.
The Bayes factors obtained by this procedure are shown in Table~\ref{tab:bayes_factor}.
We checked that our results are consistent with the Akaike and Bayesian information criteria.

\begin{table}
	\centering
	\begin{ruledtabular}
		\begin{tabular}{cccc}
		\textbf{Model}  & \textbf{$\Lambda$CDM} & \textbf{HD}  & \textbf{$w$CDM}\tabularnewline
		\hline 
		$\ln B$  & $0.00 $  & $0.88$  & $-2.63$\tabularnewline
		\end{tabular}
	\end{ruledtabular}
	\protect\caption{\label{tab:bayes_factor} Maximum Likelihood estimate of the logarithm of the Bayes 
	factor $\ln B(M)$ with respect to \lcdm{}. 
	The comparison between \lcdm{} and the HD model is inconclusive, whereas $w$CDM is disfavored with regard to a HD cosmology.}
\end{table}

The comparison of a HD cosmology and $w$CDM reveals a strong evidence for the former. There also is positive evidence for \lcdm{} over $w$CDM.
The comparison between the HD scenario and \lcdm{}, however, is inconclusive in the light of present data. The small difference between these 
two models should not be interpreted as a statistical preference for a HD cosmology since the evidence, and thus
the Bayes factor, depends on the prior volume \cite{Heavens:2017afc, Heavens:2017hkr}, which 
we computed for each of the models using the boundaries in Table~\ref{tab:priors}. If we had e.g. chosen 
a wider $w_0$ prior, the preference for the HD model over \lcdm{} would have decreased by 
$\ln \left(\Delta w_{\text{new}}/\Delta w_{\text{old}}\right)$. Given the current data sets,
we cannot establish a clear preference for a \lcdm{} or a HD cosmology. This immediately 
raises the question whether future surveys will be able to distinguish the two scenarios.

\section{Future constraints} \label{sec:forecast} 

In this section we forecast how well the HD model can be distinguished from \lcdm{} and $w\rm{CDM}$ using future 
galaxy redshift surveys. We choose three reference surveys which are planned to operate within the next decades.

As a  first reference survey we will consider a DESI-like observer. DESI\footnote{http://desi.lbl.gov/} 
is a ground-based experiment  scheduled to start in 2018. It will study the large-scale structure formation in the Universe 
using 30 million spectroscopic redshifts and positions from galaxies and quasars \cite{desi_collaboration_desi_2016-1,desi_collaboration_desi_2016,levi_desi_2013}.
For the forecast presented in this paper we use the specifications for the Emission Line Galaxies, as
explained in Ref.~\cite{desi_collaboration_desi_2016-1}. The geometry and redshift binning 
specifications, as well as the galaxy number density and bias, can be found in Refs.~\cite{desi_collaboration_desi_2016-1,CASAS201773}.

Our second reference survey will be a Euclid-like galaxy redshift survey \cite{amendola_cosmology_2013, laureijs_euclid_2011}. 
Euclid\footnote{http://www.euclid-ec.org/} is a European Space Agency mission scheduled for launch in 2020. It will 
measure about 2 billion photometric galaxy images and 100 million spectroscopic redshifts, providing a detailed 
description of structure formation up to redshift $z\sim2$. To perform our forecasts we use the survey 
parameters for a Euclid-like  mission, adapted from the Euclid Redbook specifications \cite{laureijs_euclid_2011} and from 
Ref.~\cite{Amendola:2016saw}.

The third reference survey will be a SKA2-like galaxy survey. The Square Kilometer Array 
(SKA)\footnote{https://www.skatelescope.org/} is an array of radiotelescopes around the globe to be built 
in two phases SKA1 and SKA2. Here we will use the most futuristic SKA2 stage, which is scheduled to start operating in 2030 
\cite{yahya_cosmological_2015,santos_hi_2015,raccanelli_measuring_2015,bull_measuring_2015}. 
The specifications for our forecast, such as geometry, bias and number density, are taken from Refs.~\cite{santos_hi_2015,harrison_ska_2016}.

\subsection{Fisher analysis}

To forecast the outcome of the aforementioned galaxy surveys, we use the Fisher matrix 
formalism \cite{tegmark_measuring_1998,seo_improved_2007,seo_baryonic_2005},
which is a fast way of approximating the curvature of the likelihood assuming that it is Gaussian on the parameters 
around a fiducial point. We apply this formalism to two different probes, namely Galaxy Clustering (GC) and Weak Lensing (WL), which are the 
main cosmological observables for next-generation galaxy surveys. We assume formulations of the 
likelihood which are valid in the linear regime and adapt them to partially account for the mildly non-linear 
effects appearing in cosmological structure formation. We also neglect cross-correlations among GC and WL, which is
a conservative and rather pessimistic approach. This corresponds approximately to the Fisher matrix forecasting recipe specified
in the Euclid Redbook \cite{laureijs_euclid_2011}.

\subsubsection{Galaxy Clustering}

The main observable for Galaxy Clustering is the galaxy power spectrum $P_{\rm obs}$, which is the 
Fourier transform of the two-point correlation function of galaxy number counts in redshift space. 
The galaxy power spectrum follows the power spectrum of the underlying dark matter distribution $P(k)$ up to a bias factor $b(z)$.
In this work we assume this bias factor to be local and scale-independent. Note that $P_{\rm obs}$ 
depends not only on the dark matter distribution but also on additional effects coming from the mapping between redshift 
space and real space, such as redshift space distortions
or the pairwise peculiar velocity dispersion of galaxies, the so-called Finger-of-God effect (FoG). 
Neglecting further relativistic and non-linear corrections, we follow Ref.~\cite{seo_improved_2007} and 
write the observed power spectrum as
\beq
\label{eq:observed-Pk}
\Pobs (k,\mu,z)= \frac{D_{A,f}^{2}(z)H(z)}{D_{A}^{2}(z)H_{f}(z)} B^2(z) e^{-k^{2}\mu^{2}\sigma_{tot}^2}P(k,z)\,,
\eeq
with
\beq
  \sigma_{tot}^2=\sigma_{r}^{2}+\sigma_{v}^{2}\,,\hspace{5mm}
  B(z) = b(z) (1+\beta_{d}(z)\mu^{2})\, .
\eeq
Here  $\mu\equiv\cos\varphi$, with $\varphi$ the angle between the line of sight and the 3D-wavevector
$\vec{k}$. The subscript $f$ denotes the fiducial value
of each quantity,  $D_{A}(z)$ is the angular diameter distance, $H(z)$ the Hubble function and
 $\beta_{d}(z)\equiv f(z)/b(z)$, with $f\equiv d \ln G/d\ln a$ the linear growth rate 
 of matter perturbations. In the exponential factor, we have a damping term
 $\sigma^2_{r}+\sigma_v^2$, with $\sigma_r$ the error induced by spectroscopic redshift 
 measurements and $\sigma_{v}$ the one associated to the FoG effect. We marginalize over this last
parameter \cite{bull_extending_2015} and take a fiducial value $\sigma_{v} = 300$ km/s
compatible with the estimates in Ref.~\cite{de_la_torre_modelling_2012}. For more
details on the meaning and importance of the terms in Eq.~\eqref{eq:observed-Pk}, the reader is referred to 
Refs.~\cite{CASAS201773,seo_improved_2007, amendola_testing_2012}.

Assuming a Gaussian data covariance matrix, we can write the 
Fisher matrix for the galaxy power spectrum as \citep{seo_improved_2007, amendola_testing_2012} 
\beq
  F_{ij}=\frac{V_{\rm survey}}{8\pi^{2}}\int_{-1}^{+1}\dd{\mu}\int_{k_{\rm min}}^{k_{\rm max}}\dd{k}
  \left( \pdv{D}{\theta_i} D^{-1} \pdv{D}{\theta_j} D^{-1} \right)\,,
\eeq
with
\beq
 D = D(k, \mu, z) = \Pobs(k, \mu, z) + n(z)^{-1} \,.
\eeq
Here $V_{\rm survey}$
is the volume covered by the survey and contained in a redshift slice
$\Delta z$ and $n(z)$ is the galaxy number density as a function of
redshift.
The largest scales we take into account correspond to the minimum wavenumber $k_{\rm min}=0.0079\textrm{h/Mpc}$. 
The upper limit $k_{\rm max}$ depends on the specifications of the survey and on the modeling of non-linear scales.

\subsubsection{Weak Lensing}

Another important observable in future galaxy redshift surveys is the cosmic shear, which measures distortions in the 
ellipticities of galaxy images due to the light propagation in the Universe. For a comprehensive review, see 
Ref.~\cite{Bartelmann:1999yn}. Under the assumption of small gravitational potentials and large separations, the 
cosmic shear measurements can be linked to the matter power spectrum, giving access to the cosmological parameters.
The cosmic shear at a redshift bin $i$ is correlated with the cosmic shear at another redshift bin $j$. The power 
spectrum of the cosmic shear can therefore be written as a matrix with indices $i, j$, namely
\beq
\label{def_shear}
C_{ij}(\ell)=\frac{9}{4}\int_{0}^{\infty}\dd{z}\frac{W_{i}(z)W_{j}(z)H^{3}(z)\Omega_{m}^{2}(z)}{(1+z)^{4}} P_{m} \,,
\eeq
with $P_{m}$ evaluated at the scale $\ell/r(z)$ and $r(z)$ the comoving distance. In this expression $W(z)$ is a window function given by the photometric redshift distribution function 
and the galaxy number density distribution $n(z)$. For additional details on the WL formulas see, for instance, Ref.~\cite{CASAS201773}.
Finally, we can write the WL Fisher Matrix as a sum over all multipoles correlating the signal at all 
redshift bins \citep{tegmark_measuring_1998}
\begin{equation}
F_{\alpha\beta}=f_{\rm sky}\sum_{\ell}^{\lmax}\frac{(2\ell+1)\Delta\ell}{2}
  \tr \left( \pdv{C}{\theta_\alpha} \textrm{Cov}^{-1} \pdv{C}{\theta_\beta} \textrm{Cov}^{-1} \right) \, . \label{eq:FisherSum-WL}
\end{equation} 
The prefactor $f_{\rm sky}$ is the fraction of the sky covered by the survey. The high-multipole cutoff $\lmax$ encodes
our ignorance of clustering, systematics and baryon physics on small
scales. In this work we choose $\lmax = 5000$. The quantity 
\begin{equation}
\textrm{Cov}_{ij}(\ell)=C_{ij}(\ell)+\delta_{ij}\gamma_{\rm int}^{2}\tilde{N}(n_{\theta}, \mathcal{N}_{bin})_{i}^{-1}(\ell)
\end{equation}
denotes the covariance matrix of the shear power spectrum, with $\gamma_{\rm int}$ the 
intrinsic galaxy ellipticity and $\tilde{N}_{i}^{-1}$ a 
shot noise term for each redshift bin $i$. This shot noise term depends on the total number of galaxies per
$\text{arcmin}^2$, $n_{\theta}$, and on the total number of redshift bins, $\mathcal{N}_{bin}$
(for details see Ref.~\cite{CASAS201773}).

\subsection{Fiducial values and numerical forecast parameters}

\begin{table}
	\begin{ruledtabular}
		\begin{tabular}{ l  r r }
			Parameter & \lcdm{} & HD \\
			\hline
			$\Omega_{b}h^2$ & $0.0223 \pm 0.0002$ & $0.0223 \pm 0.0002$\\
			$\Omega_{cdm}h^2$ & $0.118 \pm 0.001$ & $0.118 \pm 0.001$ \\
			$h$ & $0.682 \pm 0.007$ & $0.679 \pm 0.006$\\
			$\ln(10^{10}A_s)$ & $3.08 \pm 0.03$ & $3.07 \pm 0.03$\\
			$n_s$  & $0.971 \pm 0.005$ & $(0.966 \pm 0.003)$ \\
			$w_0$  & $(-1)$ & $-0.997 \pm 0.003$ \\
			$\Sigma m_\nu$ & $0.08 \pm 0.08$ & $0.06 \pm 0.06$
		\end{tabular}
	\end{ruledtabular}
	\caption{\label{tab:fiducialvals}Fiducial values used in the Fisher analysis together with their $1\sigma$ errors, as 
	obtained from the MCMC analysis in Section \ref{sec:data1}. The 
	values in brackets are obtained by means of the consistency relations \eqref{nswcons}-\eqref{rascons} 
	in the HD model or by a theoretical constraint in the $\Lambda$CDM case. The value for $n_s$ is only 
	relevant for the \lcdm{} scenario as in the HD case this value  follows automatically from the $w_0$ one.}
\end{table}

To explore the effect of future data sets on the cosmological parameters space we first compare the HD model to a \lcdm{} cosmology. 
For \textit{each} of these two models we take the corresponding MCMC central values as fiducial values for the forecast.
Second, we compare the HD scenario to a $w$CDM cosmology mimicking the dark-energy evolution in the Higgs-Dilaton model
but without any additional constraint on the initial power spectra. In order to emphasize the impact of the 
consistency relations on the errors, we take the central values of the \textit{Higgs-Dilaton} MCMC \textit{run} as the fiducial values 
for \emph{both} the HD and $w$CDM forecasts. For completeness, we list these values in Table~\ref{tab:fiducialvals}.

In each of the two aforementioned comparisons, we perform a Fisher Matrix forecast for both GC and WL observables.  
To compute the derivatives needed to obtain the Fisher matrices, we use
the modified version of CLASS discussed in Section \ref{sec:dataMCMCanalysis} and vary the relevant cosmological 
parameters by $\pm1\%$.  While the HD cosmology puts strong constraints on the running of the spectral tilt and the 
tensor-to-scalar ratio we do not vary them here, as they will not be directly measured by the surveys 
under consideration. The same applies to the optical depth $\tau_{reio}$, which we fix to its mean MCMC value. 

In the next section we will show DESI-like, Euclid-like and SKA2-like forecasts, with different combinations of observational probes.
In this work we define a DESI-like probe as a a simplified GC probe accounting only for linear 
scales up to $k_{\rm max}=0.15$h/Mpc but no additional non-linear corrections at baryon acoustic oscillation 
scales.
A Euclid-like probe should be understood as the same GC probe mentioned above, but 
combined with a WL probe up to a maximum multipole of $\ell_{\mathrm{max}} = 5000$, using a non-linear matter power spectrum
and neglecting intrinsic alignment corrections.
For the SKA2-like forecast, we will assume a better knowledge of the non-linear effects and 
use a non-linear matter power spectrum  up to $k_{\rm max}=0.5$ h/Mpc for GC, while for WL we will use the same maximum multipole
as in the Euclid-like case. 
The real constraining power 
of future galaxy surveys will most probably be somewhere in between the last two cases, once a better 
modeling of non-linear scales is taken into account and more probe combinations are added into the analysis.
For GC we use the latest Halofit \cite{Takahashi:2012em,Bird:2011rb} version available 
in the CLASS code \cite{lesgourgues:class} for the non-linear corrections to the matter power spectrum.
This is justified since both the HD and the $w$CDM scenario have the same first-order matter perturbations as \lcdm{}.
We combine all forecasts with the results from the MCMC run in Section \ref{sec:MCMCresults} by adding the associated Fisher matrices. In doing this we neglect cross-correlations between
CMB lensing and weak gravitational lensing, as well as cross-correlations between galaxy lensing and galaxy clustering,
as was studied for the case of SKA in Ref.~\cite{Kirk:2015xqa}.

\subsection{Results}

\begin{figure*}
 \centering
 \includegraphics[width=0.9\textwidth]{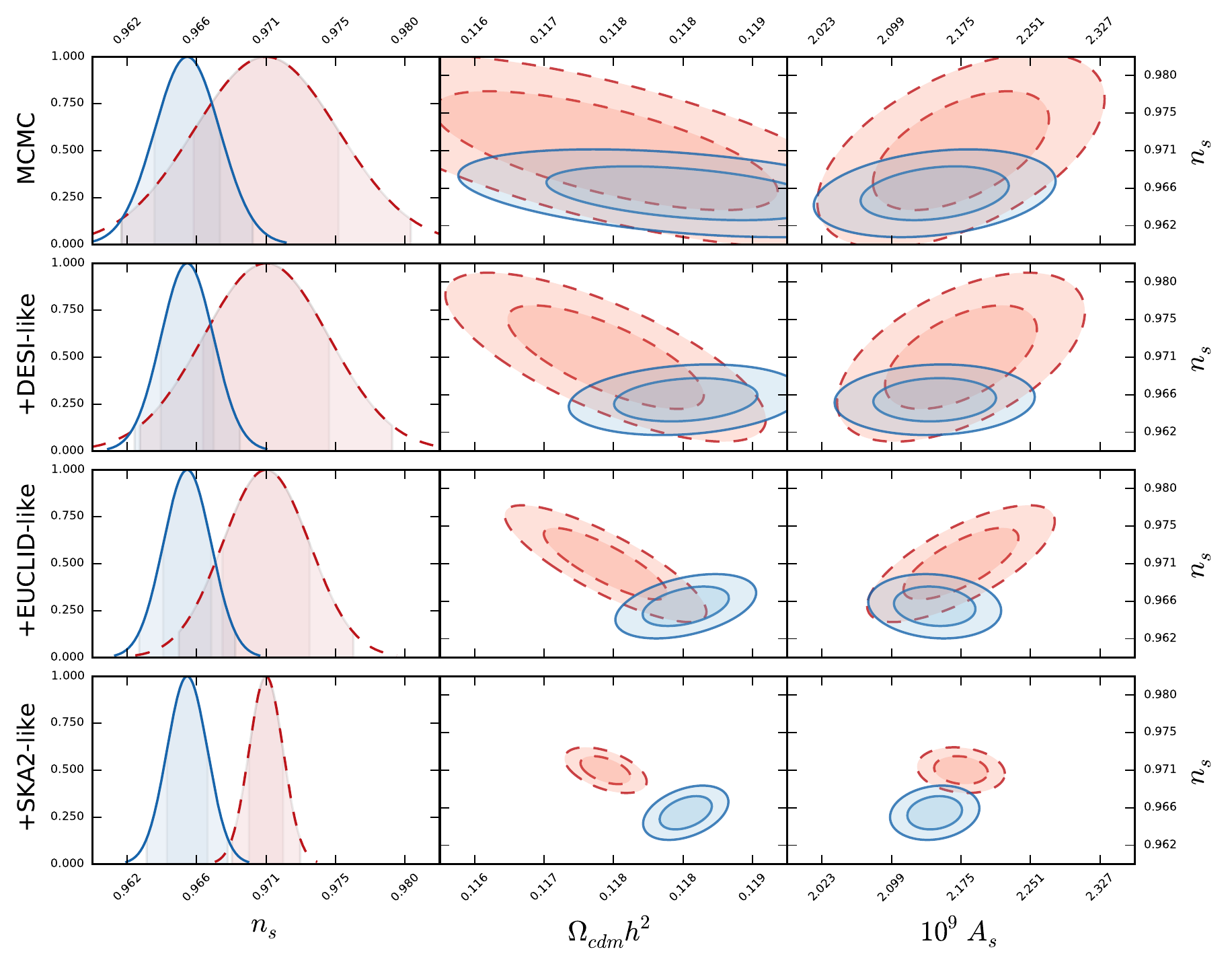}
 \caption{\label{fig:ellipses_comparisson} Comparison of a \lcdm{} (red, dashed) and a HD cosmology (blue, solid). The leftmost column
 	shows the fully marginalized 1D Gaussian probability distribution for the $n_s$ parameter. The second and third columns
 	show the $1\sigma$ and $2\sigma$ 2D Gaussian error contours. Each model is centered on the fiducial values obtained from \textit{its own} MCMC run.
  We plot i) the constraints from the MCMC run in a Fisher approximation, and add to this ii) the constraints for 
  a DESI-like mission considering GC on linear scales only iii) the constraints for a Euclid-like mission, combining GC on linear scales and WL non-linear 
  and iv) the combination of constraints for an SKA2-like survey, with both GC and WL non-linear. All remaining 
  parameters are marginalized over. A HD cosmology can be ruled out by a future measurement of large spectral tilt.}
\end{figure*}

\begin{figure*}
 \centering
 \includegraphics[width=0.9\textwidth]{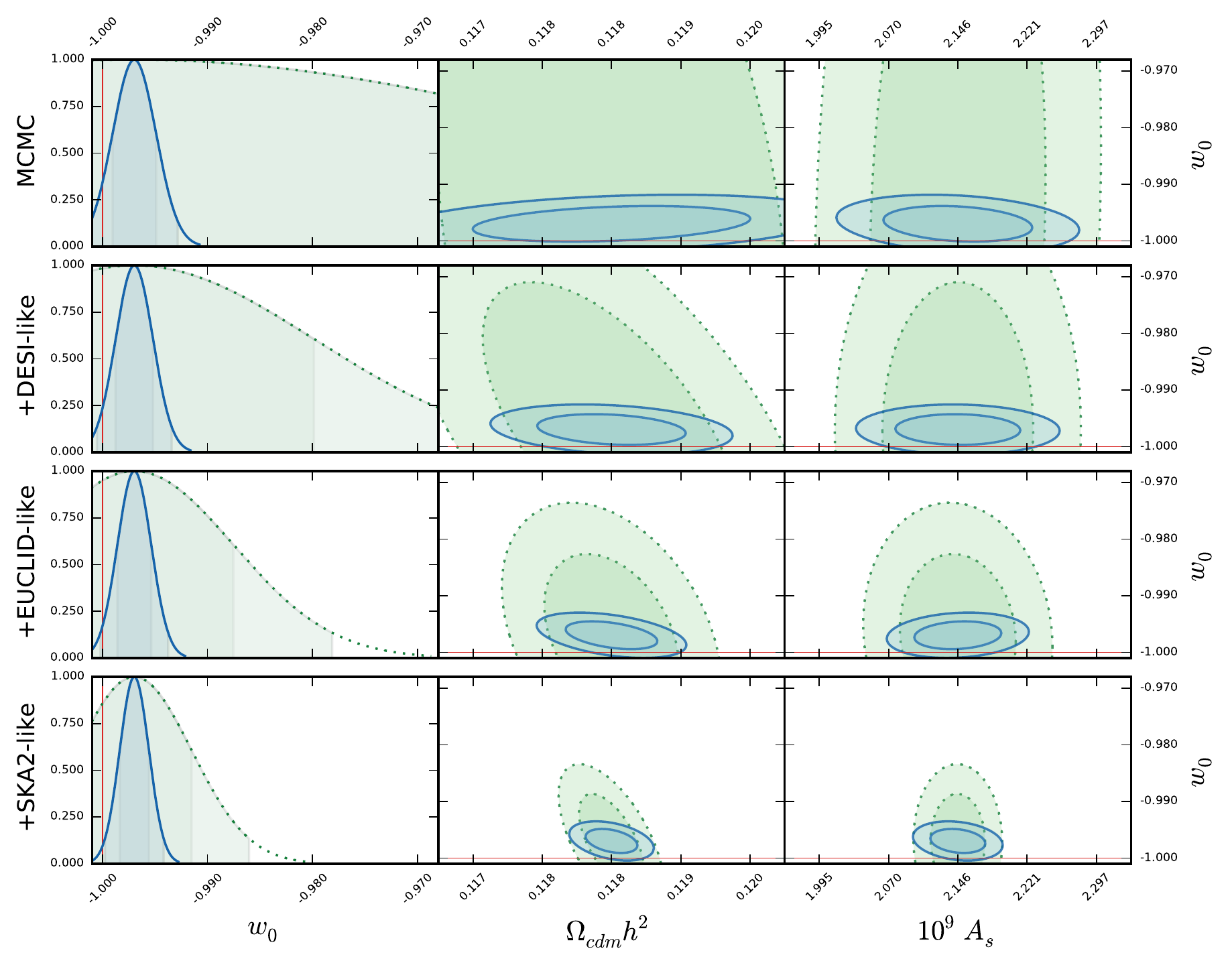}
 \caption{\label{fig:ellipses_consistency_rel} Comparison of a HD cosmology (blue, solid) and a $w$CDM model (green, dotted) for 
 the same observables and data as in  Fig.~\ref{fig:ellipses_comparisson}.  Note that the $w$CDM case is centered on the 
 \textit{Higgs-Dilaton fiducial values} (for example the MCMC mean value for $w_0$ in $w$CDM is $w_0 = -0.939\pm0.061$) to facilitate 
 the comparison with the HD cosmology and to illustrate the impact of the consistency relations on parameter uncertainties, especially on the 
 estimation of the present dark-energy equation-of-state parameter $w_0$. A HD cosmology can be ruled out by a measurement 
 of a large $w_0$ or if phantom dark energy is preferred by future data. For a futuristic SKA2-like survey, the cosmological 
 constant scenario could be ruled out at the $2\sigma$ level.}
\end{figure*}

In this section we present the results of our forecast analysis. We compute $1\sigma$ and $2\sigma$ error ellipses 
for the parameters $\Omega_{cdm}h^2$, $10^9 A_s$, $n_s$ and $w_0$ in the different scenarios and marginalize 
over all the remaining parameters.  Additionally we present histograms for the dark energy
equation-of-state today and the spectral tilt. 

In Fig.~\ref{fig:ellipses_comparisson} we compare the HD model to a standard \lcdm{} scenario. This 
comparison allows us to determine how well future data will be able to discriminate between two competing 
models that are currently close in parameter space and have a similar statistical significance. As argued in Section 
\ref{sec:MCMCresults}, for present cosmological data the HD parameter space is just a restriction of the \lcdm{} one. 
Note, however, that in the HD scenario the spectral tilt is bounded from above for a theoretically well-motivated 
number of $e$-folds. In particular, for the mean value $N_*=60.7$, we 
have $n_s\lesssim 0.967$, cf.~Eq.~\eqref{eq:bound2}.
This is a first prediction of the HD model that sets it apart from the standard cosmological scenario.
The upper bound on $n_s$ has some interesting consequences for the outcome of future 
cosmological surveys. Indeed, if the preferred value for $n_s$ moves to larger values in the future, even a 
pure GC DESI-like probe will be able to find clear differences between a HD cosmology and a \lcdm{} one.
If the current mean values for these two models are maintained in the presence of new data sets a combination of 
a Euclid-like or SKA2-like probe with present datasets translates into a $\Lambda$CDM value of $n_s$ 
which is more than $3\sigma$ away from the HD one, when analyzed from the point of view of a HD cosmology. A Bayesian inference approach will potentially
be able to discriminate between these two well-separated cases.
In Table \ref{tab:forecast_errors} we list the estimated errors on the interesting parameters for a Euclid-like 
probe plus current cosmological data. This shows that the constraints on $n_s$ will improve by a factor 2 in HD cosmology, mostly due
to the effect of the consistency relations.

\begin{table}
	\centering
	\begin{ruledtabular}
		\begin{tabular}{cccc}
			  & \textbf{\lcdm{}} & \textbf{HD}  & \textbf{$w$CDM}\tabularnewline
			\hline 
			$\Omega_{cdm}h^2$  & 0.40\%  & 0.28\%  & 0.43\% \tabularnewline
			$10^9 A_s$  & 1.9\%  & 1.4\%  & 1.9\% \tabularnewline
			$n_s$  & 0.29\%  & 0.16\%  & 0.31\%\tabularnewline
			$w_0$  & -  & 0.16\%  & 0.95\% \tabularnewline
		\end{tabular}
	\end{ruledtabular}
	\protect\caption{\label{tab:forecast_errors} $1\sigma$ constraints from a Euclid-like probe combined with the MCMC covariance 
	matrices obtained from present data, for the different models considered in this work. The constraining power of the consistency relations 
	manifest as a reduction of the error on $n_s$ by a factor 2 and on a reduction of the error on $w_0$ by a factor 6 in 
	the HD scenario. For the DESI-like survey we find the same trend, but with all relative errors roughly 50 \% larger, while for the SKA2-like observation, the constraints improve roughly by a factor 2, consistent with the elliptical contours in Figs.~\ref{fig:ellipses_comparisson} and \ref{fig:ellipses_consistency_rel}.}
\end{table}

The results presented in Fig.~\ref{fig:ellipses_consistency_rel} allow to estimate the constraining power of the consistency relations. 
 In this figure, we display the error contours for the HD scenario and for a 
 $w$CDM cosmology \textit{centered at the} HD \textit{central values} in order to facilitate the comparison. 
For present data sets, the error contours in the HD model are well inside the $w$CDM contours, showing that
the constraining power of Planck data on the spectral tilt is enough to restrain the HD model in a 
way far beyond the limits of a model without consistency relations. The HD model is capable of predicting 
the complete background evolution of the Universe and therefore a precise value for the dark energy equation-of-state parameter. 
This is a second clear prediction that is easy to test with future observations. Indeed, when analyzed with present 
data, the $w$CDM model provides a mean value $w_0=-0.939 \pm 0.061$. 
If future data would favor this value of the dark energy equation of state, the HD scenario would be strongly disfavored. 
To accommodate a value $w_0 \gtrsim -0.98$ and at the same time agree with measurements of $n_s$, 
one would need a number of $e$-folds of inflation in strong tension with theoretical estimates. 
In the same way, a measurement of a  dark-energy equation-of-state parameter very close to that of a cosmological constant 
will also disfavor the HD model by roughly $2 \sigma$ in the more futuristic case.

\begin{figure}
	\includegraphics[width=0.45\textwidth]{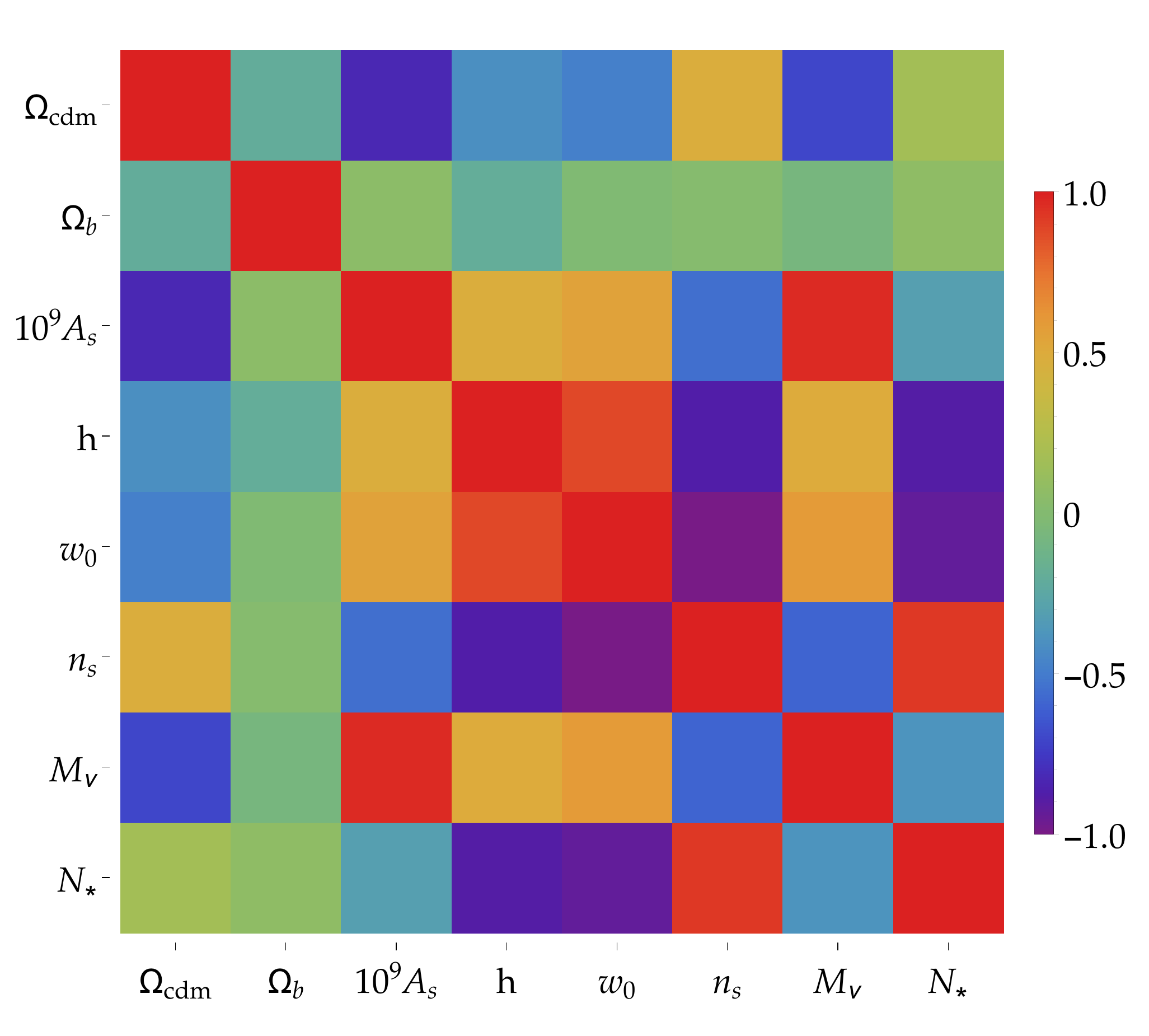}
	\caption{\label{fig:corrmat_HD_Euc} Correlation matrix for the HD model forecast for a Euclid-like probe. 
	The $+1$ and $-1$ limits stand respectively for \textit{totally correlated} and
	 \textit{totally anticorrelated}.}
\end{figure}

The error contours in Figs.~\ref{fig:ellipses_comparisson} and \ref{fig:ellipses_consistency_rel} also illustrate
how the constraints on standard parameters get affected by the
consistency relations. For instance, in  Fig.~\ref{fig:ellipses_comparisson} one can observe that the 
degeneracy direction between $\Omega_{cdm}h^2$ and $n_s$ as well as the degeneracy directions
between $10^9 A_s$ and $n_s$ are rotated in the HD model (blue solid ellipses) as compared to the \lcdm{} case (red dashed ellipses).
This can be understood in terms of the correlation coefficients of the corresponding covariance matrices, 
see Fig.~\ref{fig:corrmat_HD_Euc}.
Due to the strong correlation of $n_s$ and $w_0$ with $N_*$, of the order of $+1$ and $-1$ respectively,
other correlations present in standard cosmology -- for example the negative correlation between $\Omega_{cdm}$ and $n_s$
and the positive correlation between $h$ and $w_0$ -- get sufficiently altered, such that in the 
HD cosmology one observes a positive correlation between $n_s$ and $\Omega_{cdm}$.
This breaks degeneracies in parameter space and helps to better constrain other cosmological parameters, 
although only slightly for the case of $\Omega_{cdm}h^2$ and $10^9 A_s$, cf.~Table \ref{tab:forecast_errors}.

The most important feature of the HD model is that it singles out a curve in the parameter space spanned by $w_0, r, n_s, \alpha_s$.
A precise measurement of two or more of these parameters provides a consistency check of the HD model. If the parameters inferred from data assuming a model different from HD cosmology do not
fall onto the HD curve, the HD model will be challenged, and eventually the Bayesian evidence would favor other models.

We finish this section with a disclaimer. Although the Fisher Matrix technique allows for a quick estimate of the constraining power 
of future surveys, it is rooted on the assumption of a 
Gaussian approximation for the likelihood, which certainly fails for sharp restrictions such as the absence of
negative neutrino masses or the absence of phantom behavior with $w_0<-1$ assumed in this paper. 
For this reason, it would be convenient to eventually perform a more robust forecast by 
sampling the observational likelihoods via an MCMC approach,
 which, in our case, should reflect the consistency relations our model imposes on the parameter space. 
In the context of this analysis it would also be interesting to account for the effect of the non-flat priors suggested in 
Ref.~\cite{Hannestad:2017ypp}. These could potentially lead to mean values of $w_0$ in the HD case even further 
away from $w_0=-1$.

\section{Conclusion}\label{sec:conclusions} 

The Higgs-Dilaton model is a scale-invariant extension of the Standard Model non-minimally coupled to unimodular gravity 
and containing just an additional degree of freedom on top of the Standard Model particle content. This minimalistic 
framework allows for a successful inflationary stage followed by a standard hot Big Bang era and a late-time dark-energy
domination period. The inflationary and dark energy dominated eras turn out to be strongly related in the 
Higgs-Dilaton scenario. In particular, the model predicts a set of measurable consistency
relations between the inflationary observables and the dark-energy equation-of-state parameter. We presented 
an alternative derivation of these consistency relations that highlights the connections and differences 
with $\alpha$-attractor scenarios \cite{Ferrara:2013rsa,Kallosh:2013yoa,Galante:2014ifa} and allows for a straightforward
generalization to the general scale-invariant scenarios considered  in Ref.~\cite{Karananas:2016kyt}. We studied
the impact of the Higgs-Dilaton consistency relations on the analysis of present data sets and on the results of future galaxy surveys. To 
this end, we compared the Higgs-Dilaton model to a standard $\Lambda$CDM cosmology and to a $w$CDM scenario mimicking 
the dark-energy evolution of the Higgs-Dilaton model but without any additional  constraints on the primordial 
power spectra. In the light of present data sets our results show that the Higgs-Dilaton model is preferred 
with respect to $w$CDM but still statistically indistinguishable from a $\Lambda$CDM cosmology. To estimate the 
discriminating power of future galaxy surveys, we used a Fisher Matrix approach to perform a forecast for
the different scenarios, both for galaxy clustering 
and weak lensing probes. Assuming that the best fit values in these models remain comparable to the present values, we showed 
that both Euclid- and SKA2-like missions will be able to discriminate a Higgs-Dilaton cosmology from $\Lambda$CDM 
and $w$CDM. In particular, the Higgs-Dilaton model singles out a curve in the multiparameter space spanned by the spectral tilt 
of curvature perturbations, its running, the tensor-to-scalar ratio and the dark-energy equation-of-state parameter. A precise 
measurement of two or more of these parameters will provide a check of the HD model. On top of that, the strong correlation 
among them breaks degeneracies in parameter space by modifying other parameters' correlations without significantly 
altering their mean value with respect to \lcdm{}.

\begin{acknowledgments}
We acknowledge support from the DFG through the project TRR33 ``The Dark Universe''. MP acknowledges 
support by the state of Baden-Württemberg through bwHPC.

\end{acknowledgments}

\bibliography{references}

\end{document}